\documentclass[journal]{IEEEtran}

\usepackage{amsmath,amsfonts}
\usepackage{algorithm}
\usepackage{algpseudocode}
\usepackage{algorithmicx}
\usepackage{array}
\usepackage{textcomp}
\usepackage{stfloats}
\usepackage{url}
\usepackage{verbatim}
\usepackage{graphicx}
\usepackage{subfigure}
\usepackage{multirow}
\usepackage{cite}
\usepackage{xcolor}
\usepackage{makecell}
\usepackage{tcolorbox}
\tcbuselibrary{breakable}
\usepackage{enumitem}
\setenumerate[1]{itemsep=0pt,partopsep=0pt,parsep=\parskip,topsep=0pt}
\setitemize[1]{itemsep=0pt,partopsep=0pt,parsep=\parskip,topsep=0pt}
\setdescription{itemsep=0pt,partopsep=0pt,parsep=\parskip,topsep=0pt}

\newcounter{ccounter}
\newcounter{jcounter}
\newcommand{\cc}{
\#\theccounter
\addtocounter{ccounter}{1}}

\newcommand{\jc}{
\#\thejcounter
\addtocounter{jcounter}{1}}

\begin{document}
%
\title{Black-Box Performance Evaluation of Elastic Block Storage: Contract, Rate-Limiting Model, and Software Exploration}
%
%
%

\author{Yingjia~Wang,~\IEEEmembership{Student Member,~IEEE,}
        and~Ming-Chang~Yang,~\IEEEmembership{Member,~IEEE}
\thanks{
This paper is extended from~\cite{wang2025unwritten}, which was published in the 62nd ACM/IEEE The Chips to Systems Conference (DAC), San Francisco, CA, USA, June 22-25, 2025.
Yingjia Wang and Ming-Chang Yang are both with the Department of Computer Science and Engineering, The Chinese University of Hong Kong, Sha Tin, NT, Hong Kong.}
}

%
%

\markboth{IEEE Transactions}%
{Wang \MakeLowercase{\textit{et al.}}: Black-Box Performance Evaluation of Elastic Block Storage}
%



\maketitle

\begin{abstract}

Elastic block storage (EBS) with the storage-compute disaggregated architecture is a key component in modern cloud infrastructure.
EBS offers users storage resources in the form of elastic solid-state drives (ESSDs).
Nonetheless, despite recent efforts that have documented EBS architectures from the provider's perspective, how ESSDs perform differently from local SSDs and how host software should adapt accordingly have not been sufficiently studied.

In this paper, we conduct a user-centric, black-box performance characterization of ESSDs from Amazon AWS and Alibaba Cloud.
We make three main contributions: (1) an ESSD contract that presents four behavioral observations and five actionable implications for software adaptation, (2) a refined I/O rate-limiting model combining bandwidth-IOPS dual limiting and fine-grained token refilling to suppress latency spikes, and (3) a case study on RocksDB that derives four guidelines on cache management, I/O regulation, storage budget utilization, and compression algorithms.
Collectively, we hope these contributions can serve as a practical reference for EBS users to understand and exploit the distinctive performance properties of ESSDs.

\end{abstract}


%

\section{Introduction} \label{sec:intro}

\textit{Elastic block storage (EBS)} serves as a foundational component in modern cloud infrastructures, meeting the high-performance storage requirements of diverse cloud applications and services~\cite{miao2022luna,li2023depth,zhang2024s,wang2024ransom,shu2024burstable,wu2025hey}.
EBS commonly employs a disaggregated architecture that separates compute and storage into distinct clusters, interconnected via high-speed data center networks.
Within this architecture, user virtual machines (VMs) run in the compute cluster, while data is persistently stored in the storage cluster.
This separation not only improves the scalability and operational efficiency of the cloud infrastructure but also provides users with flexible and pay-as-you-go pricing that enhances their cost-effectiveness.

EBS provides storage to users through \textit{elastic solid-state drives (ESSDs)}, which are virtual storage devices attached to user VMs.
From the user’s perspective, an ESSD resembles a conventional local SSD, exposing a standard block interface and allowing existing filesystems and applications to operate without modifications.
However, a key distinction of ESSDs is their ability to overcome the physical performance and capacity limits typical of local SSDs.
This enables dynamic adjustment of performance (e.g., in throughput and IOPS) and storage capacity, adapting seamlessly to evolving user needs.
Additionally, ESSDs deliver high data availability and reliability by leveraging cloud-level replication and redundancy, along with advanced capabilities including the support of encryption and snapshotting.

The expansion of cloud computing has accelerated the integration of ESSDs across a wide range of cloud services.
However, although recent provider-side studies~\cite{zhang2024s,xu2025evolving} have documented the architectures and design rationales of EBS from the provider's perspective, the performance characteristics \textit{from the users' viewpoint} have not been sufficiently evaluated.
This gap is consequential, as the existing host software is extensively optimized under assumptions derived from local SSDs, yet ESSDs may violate these assumptions.
It is particularly vital for users to understand ESSDs' unique performance features, especially when compared to well-known local SSDs, and how to adapt their software accordingly.

In this paper, we conduct a \textit{black-box performance evaluation} of the world’s two major EBS service providers: Amazon AWS and Alibaba Cloud.
Our first contribution is an ESSD \textit{contract}, structured around four \textit{observations} and five \textit{implications}.
Three of the observations capture \textit{unwritten} behaviors of ESSDs related to latency, garbage collection (GC), and access pattern; they are undocumented and challenge our conventional understanding derived from local SSDs.
The last observation experimentally validates the \textit{written} performance budget, showing that bandwidth and IOPS are deterministic and contrast with the pattern-dependent peaks of local SSDs.
Building on the observations, the implications offer five actionable recommendations to guide ESSD users in re-evaluating the architecture of their deployed software.

Our second contribution in this paper is \textit{a refined I/O rate-limiting model for cloud software to obtain satisfactory latency.}
Zhou et al.~\cite{zhou2023calcspar} first point out that it is crucial for cloud software to limit I/O rates to ESSDs in accordance with the internal rate-limiting within EBS, since exceeding the IOPS budget per second regulated by the EBS provider can lead to substantial latency spikes.
However, our experiments identify three additional factors that affect latency: bandwidth oversubscription, provider-specific token-refill intervals, and latency spikes caused by non-instantaneous synchronization between cloud-software-side and EBS-side rate-limiting. 
Accordingly, our model combines \textit{bandwidth-IOPS dual limiting} with \textit{fine-grained token refilling}, avoiding both bandwidth and IOPS budget violations while mitigating latency spikes caused by coarse-grained token refilling.

Finally, our third contribution is an application-level exploration of the contract and the rate-limiting model through a case study on RocksDB~\cite{website:rocksdb}, a widely deployed LSM-tree-based key-value store.
We derive four actionable guidelines from our exploration.
First, the high per-I/O cost of ESSDs makes cache miss minimization critical to meeting read SLAs, requiring more aggressive block cache provisioning than in local-SSD deployments.
Second, by instantiating our refined rate-limiting model into RocksDB, we reduce the maximum latency by up to 97.8\% compared with Calcspar's one-second IOPS rate-limiting model~\cite{zhou2023calcspar} and consistently maintain it at tens of milliseconds across various YCSB benchmarks~\cite{cooper2010benchmarking}.
Third, the absence of user-visible GC-related degradation suggests that cloud databases may reconsider the conservative capacity and performance headroom inherited from local-SSD deployments.
Finally, enabling aggressive compression with standard algorithms (e.g., ZSTD~\cite{website:zstd}) on ESSDs, compared to LZ4~\cite{website:lz4}, yields up to 67.2\% higher write throughput and up to 49.3\% smaller storage footprints, while read latency remains comparable on highly compressible data.
Collectively, these results demonstrate the potential of ESSD-aware optimization in the evaluated RocksDB deployment without requiring any hardware change.

In the remainder of this paper, Section~\ref{sec:bg} presents the background of (local) SSD, EBS, and ESSD.
Sections~\ref{sec:uc},~\ref{sec:model}, and~\ref{sec:cs} elaborate on the contract, the rate-limiting model, and the RocksDB case study, respectively.
Section~\ref{sec:related} introduces the related work. Section~\ref{sec:conclusion} concludes our work.

\section{Background} \label{sec:bg}

\subsection{NAND Flash-based Solid-State Drive (SSD)} \label{sec:bg_ssd}

NAND flash-based solid-state drives (SSDs) have gained a significant and growing portion of the storage market, owing to their advantages in speed, power efficiency, and physical durability.
SSDs organize storage into hierarchical tiers: channels, dies, planes, blocks, and pages. The flash die is the smallest unit for parallel operations, while the flash page represents the smallest unit that can be independently read or programmed. To improve firmware efficiency, flash blocks are aggregated into superblocks, each spanning multiple dies to maximize parallelism.

Despite these internal tiers, SSDs typically expose a simple abstraction to the host: the \textit{block interface}, which represents the storage medium as a sequence of logical blocks (e.g., 4KB) supporting random read and write operations.
Although this straightforward abstraction has played a key role in the widespread use of SSDs, it necessitates a sophisticated \textit{flash translation layer (FTL)} within the drive to accommodate the unique properties of NAND flash, such as the requirement to erase before writing.
The core functions of the FTL include address mapping and garbage collection (GC).
Address mapping involves translating logical addresses provided by the host into physical locations within the flash storage, maintained via a fine-grained (e.g., page-level) mapping table.
Garbage collection is performed at regular intervals to reclaim invalid space at the block level. 
The FTL relocates valid pages from selected blocks so they can be erased and reused.
The FTL also handles other essential management tasks, including wear-leveling, error correction coding, and bad block management.

The management overhead of the traditional FTL has become a key bottleneck for SSD performance and cost reduction. New NVMe interface standards, including Zoned Namespace (ZNS)~\cite{bjorling2021zns} and Flexible Data Placement (FDP)~\cite{sabol2023fdp}, address this by enabling host-managed or host-guided data placement, which separates data by lifetime into distinct superblocks and minimizes device-side GC. FDP achieves these benefits while preserving backward compatibility with the conventional block interface. ZNS, in contrast, requires host software adaptation for zone management but further reduces address mapping overhead and lowers cost.

\subsection{Elastic Block Storage (EBS)} \label{sec:bg_ebs}

\begin{figure}[t]
    \centering
    \includegraphics[width=0.4\textwidth]{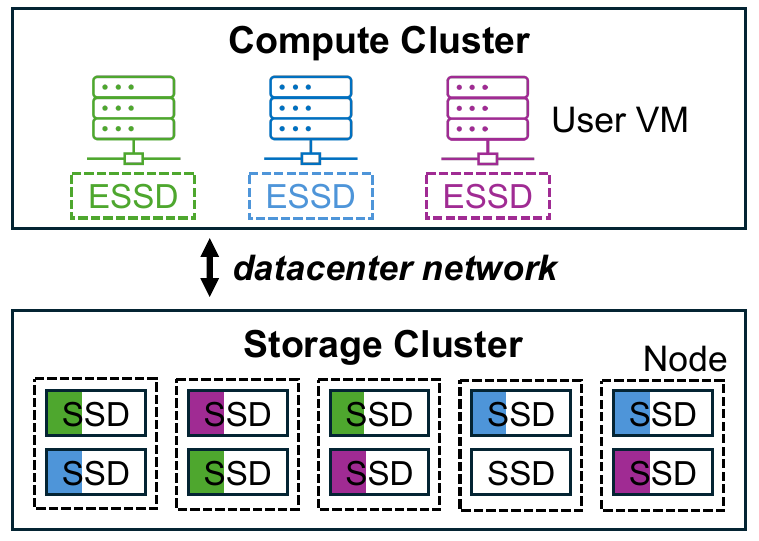}
    \caption{
    \textbf{The disaggregated storage-compute architecture of elastic block storage (EBS).}
    The ESSD is attached to the user's virtual machine (VM) located within the compute cluster.
    Its physical storage is distributed and replicated across multiple nodes and SSDs in the storage cluster.
    }
    \label{ebs}
\end{figure}

As illustrated in Figure~\ref{ebs}, \textit{elastic block storage (EBS)} commonly utilizes a disaggregated storage-compute architecture, where separate compute and storage clusters are deployed. These clusters are not physically co-located but are connected via a high-speed data center network.
The \textit{compute cluster} operates users' virtual machines (VMs) and relays their I/O requests to the storage cluster. Meanwhile, the \textit{storage cluster} is responsible for durably storing and retrieving data through a distributed filesystem.


This separation of storage and compute offers several advantages~\cite{miao2022luna,shu2024burstable,zhang2024s}.
First, both types of resources can be provisioned and billed elastically based on user workload requirements. This leads to improved resource utilization and greater cost efficiency compared to general-purpose server offerings.
Second, independent maintenance and scaling of compute and storage clusters become feasible, enhancing both maintainability and scalability.
Third, migrating application services between compute servers is more straightforward and efficient, as their state can be consistently maintained in the remote storage servers.



\subsection{Elastic Solid-State Drive (ESSD)} \label{sec:bg_essd}

EBS offers storage resources to users through \textit{elastic solid-state drives (ESSDs)}, which are virtual storage devices attached to user VMs.
These ESSDs can be created, attached, detached, or deleted without interfering with the normal operation of the user's VM.
As depicted in Figure~\ref{ebs}, each virtual ESSD maps to one or more physical SSDs.
In practice, the physical storage of an ESSD is typically distributed and replicated (e.g., using three-way replication~\cite{zhang2024s}) among different nodes and SSDs within the storage cluster. This strategy supports load balancing and enhances data availability.


From the user’s perspective, an ESSD resembles a traditional local SSD in that it also presents a block interface and allows random access within the storage space.
This design ensures full compatibility with existing software ecosystems, enabling filesystems and applications to run on ESSDs without any modification.
However, ESSDs provide several advantages over local SSDs~\cite{zhang2024s,website:awsssd,website:alissd}.
First, as a virtualized device, an ESSD is not constrained by the physical limitations of a single SSD, allowing both performance (e.g., throughput and IOPS) and capacity to be scaled elastically according to dynamic user needs.
Second, ESSDs offer high data availability and reliability by replicating data across multiple nodes, thereby avoiding single points of failure.
Third, ESSDs support advanced functionalities including encryption and snapshot capabilities.



Cloud storage providers commonly supply a range of ESSD types, designed for distinct performance tiers.
These tiers come with differing provisioned volume limits in terms of maximum throughput and IOPS, with higher performance levels generally offered at increased cost.
For example, Amazon AWS~\cite{website:awsssd} provides both general-purpose categories (such as gp2 and gp3) and provisioned-IOPS categories (including io1 and io2), among which io2 represents the highest performance option.
Similarly, Alibaba Cloud~\cite{website:alissd} offers four graduated performance levels (i.e., PL0, PL1, PL2, and PL3), where a higher number corresponds to better performance.



\begin{small}
\begin{table*}[t]
    \caption{\textbf{The configurations of evaluated storage devices (i.e., ESSDs and SSD).}
    }
    \centering
    \small
    \begin{tabular}{c|c|c|c|c|c|c}
    \hline
    & \textbf{Provider and Type} & \textbf{Max. BW (GB/s)} & \textbf{Max. IOPS (K)} & \textbf{Cap. (TB)} & \textbf{VM Type} & \textbf{Region} \\ \hline
\textbf{\textit{ESSD-AM1}} & Amazon AWS io2 (high-end) & $\sim$3.1 & $\sim$100 & 2 & m6in.xlarge & Tokyo \\ \hline
\textbf{\textit{ESSD-AM2}} & Amazon AWS gp3 (low-end) & $\sim$0.13 & $\sim$5 & 0.25 & m6in.xlarge & Tokyo \\ \hline
\textbf{\textit{ESSD-AL1}} & Alibaba Cloud PL3 (high-end) & $\sim$1.1 & $\sim$105 & 2 & ecs.g5.4xlarge & Shanghai \\  \hline
\textbf{\textit{ESSD-AL2}} & Alibaba Cloud PL0 (low-end) & $\sim$0.18 & $\sim$5 & 0.25 & ecs.g5.4xlarge & Shanghai \\ \hline
\textbf{\textit{SSD}} & Samsung 970 Pro & \makecell[c]{Seq. R/W \\$\sim$3.6/$\sim$2.6}  & \makecell[c]{Rand. R/W\\$\sim$800/$\sim$620} & 1 & / & / \\
    \hline
    \end{tabular}
    \label{essd_config}
\end{table*}
\end{small}

\section{The Contract of ESSDs} \label{sec:uc}

Despite the widespread adoption of ESSDs across various cloud platforms, their performance characteristics remain, to our knowledge, insufficiently studied.
In this section, we present a contract of ESSDs.
We first describe our experimental setups in Section~\ref{sec:uc_expsetup}, followed by a detailed presentation of the contract across Sections~\ref{sec:uc_lat} to~\ref{sec:uc_mix}.



\subsection{Experimental Setups} \label{sec:uc_expsetup}

To enhance the generalizability and thoroughness of our results, we evaluate four real-world ESSD instances from two major EBS providers: Amazon AWS~\cite{website:awsssd} and Alibaba Cloud~\cite{website:alissd}.
The ESSD types include Amazon AWS io2 (high-end), Alibaba Cloud PL3 (high-end), Amazon AWS gp3 (low-end), and Alibaba Cloud PL0 (low-end).
Detailed configurations for these ESSDs are provided in Table~\ref{essd_config}.
For comparative analysis with local SSDs, we also include the Samsung 970 Pro~\cite{website:samsung970pro}, a widely-used model in research. Its specifications are likewise summarized in Table~\ref{essd_config}.
Testing for the Samsung 970 Pro is conducted on a server configured with four 24-core/48-thread Intel(R) Xeon(R) Platinum 8160 CPUs, running a Debian 5.10.223-1 operating system with Linux kernel version 5.10.0.

We utilize the FIO benchmark tool~\cite{website:fio} in conjunction with the io\_uring engine~\cite{website:iouring} to generate workloads with varied load conditions (e.g., different I/O sizes and queue depths) and access patterns (e.g., random/sequential reads and writes).
The direct I/O mode is employed to bypass the page cache and the none scheduler is used to prevent request merging within the block layer.
Further specifics regarding workload parameters are detailed in the corresponding subsections.
In the following experiments, all metrics presented are averaged across 5 repeated runs, except for GC experiments in Section~\ref{sec:uc_gc}. 
In GC experiments, similar trends are observed across 3 repeated runs, so we depict the data of one run.

\subsection{Latency} \label{sec:uc_lat}

\begin{figure*}[!t]
    \centering
    \subfigure[Average Latency of ESSD-AM1 from Amazon AWS]{
    \includegraphics[width=0.48\textwidth]{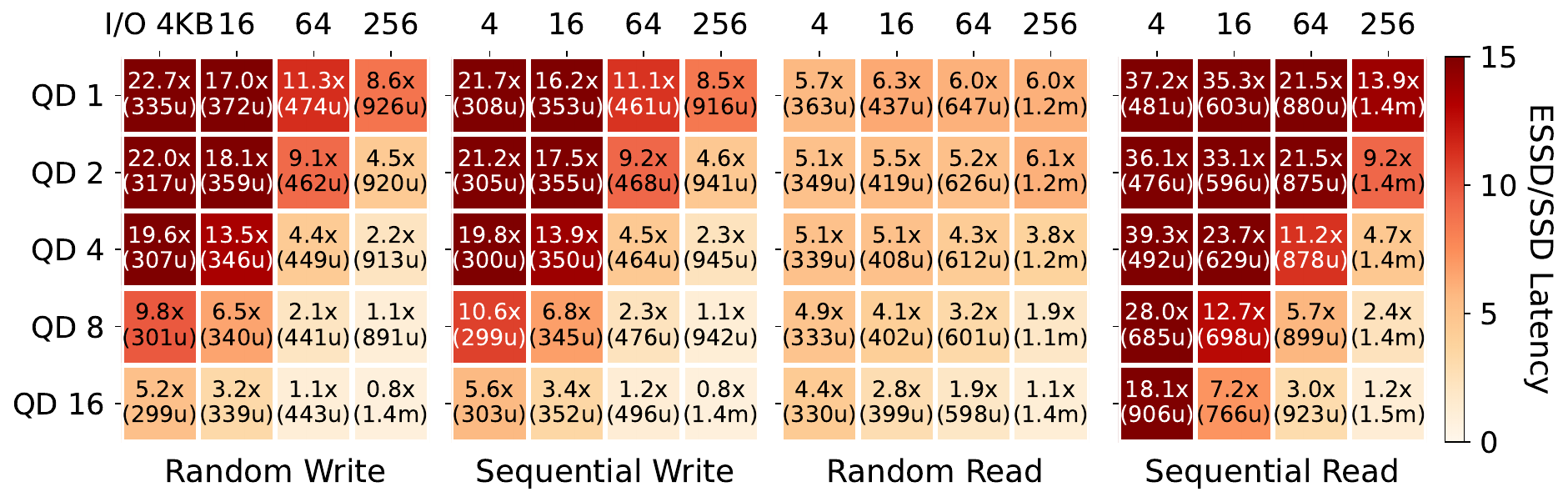}
    }
    \subfigure[P99.9 Latency of ESSD-AM1 from Amazon AWS]{
    \includegraphics[width=0.48\textwidth]{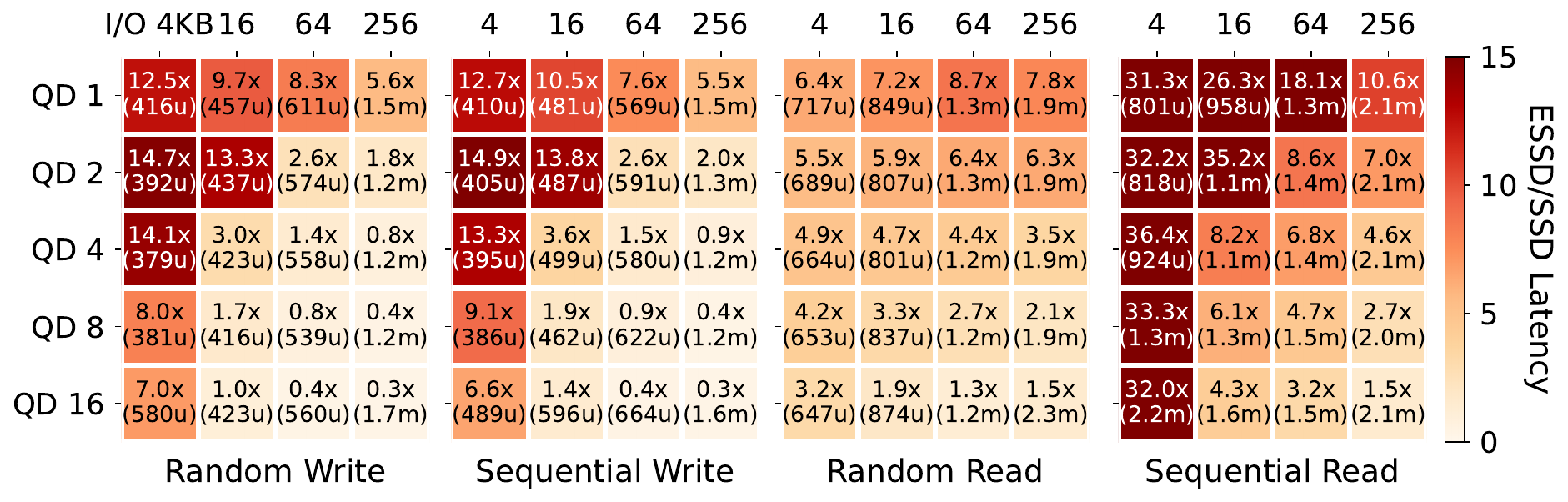}
    }
    \subfigure[Average Latency of ESSD-AL1 from Alibaba Cloud]{
    \includegraphics[width=0.48\textwidth]{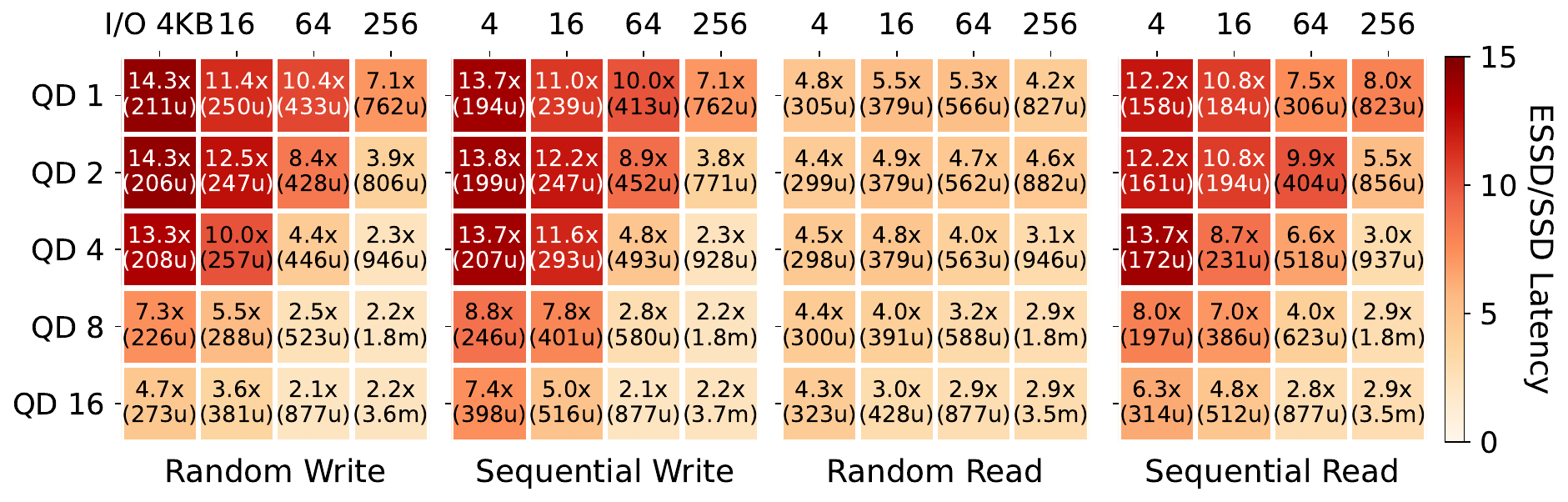}
    }
    \subfigure[P99.9 Latency of ESSD-AL1 from Alibaba Cloud]{
    \includegraphics[width=0.48\textwidth]{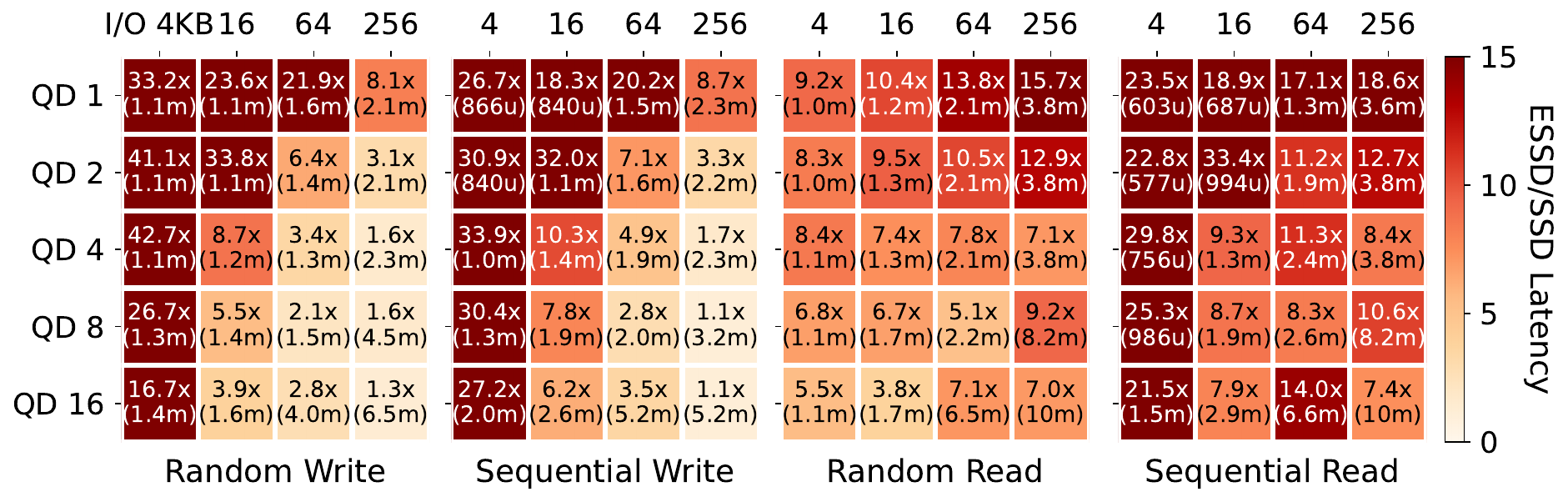}
    }
    \caption{
    \textbf{Latency of two high-end ESSDs under workloads with different access patterns (i.e., random write, sequential write, random read, sequential read), I/O sizes, and I/O queue depths (QD).}
    The top of each pixel indicates the multiples the ESSD latency is divided by the SSD latency, with smaller numbers being better.
    The bottom of each pixel indicates the value of average or P99.9 latency of the ESSD (u: $\mu$s, m: ms).}
    \label{exp_lat_high}
\end{figure*}

\vspace{.5em}
\noindent\fbox{\parbox{0.475\textwidth}{
\textbf{Observation\cc:} \emph{Even with high-end ESSDs, their latency is tens of times higher (e.g., up to 39.3$\times$ and 42.7$\times$) than that of SSD when I/Os are not well scaled up. Here, “not well scaled up” is defined as workloads with small I/O sizes and/or low I/O queue depths.}
}}
\vspace{.5em}

Ideally, ESSDs should deliver performance comparable to local SSDs. However, our evaluation reveals that even high-end ESSDs (i.e., ESSD-AM1 from Amazon AWS and ESSD-AL1 from Alibaba Cloud) exhibit significantly higher latency when I/Os are not well scaled up, defined as workloads with small I/O sizes and/or low I/O queue depths.


Figure~\ref{exp_lat_high} illustrates the average and 99.9th percentile (P99.9) latency of these two ESSDs across various access patterns (i.e., random write, sequential write, random read, and sequential read), with I/O sizes ranging from 4KB to 256KB and queue depths from 1 to 16.
Of particular interest is the ratio of ESSD latency to local SSD latency (indicated above each pixel in the figure), where lower values reflect better performance.
For clarity, we refer to this as the \textit{latency gap} in this subsection.


A clear trend emerges: the latency gap narrows as I/Os scale up in either size or queue depth, consistently across both metrics, all four access patterns, and both providers.
Under small I/O conditions, the average latency gap reaches 39.3$\times$ and 14.3$\times$ for the two ESSDs, with P99.9 gaps peaking at 36.4$\times$ and 42.7$\times$.
With 256KB I/Os, these gaps shrink substantially (e.g., average gaps of 13.9$\times$ and 8.0$\times$), and similar reductions occur with higher queue depths.


This high latency stems from two factors. First, network delays and software processing overhead in cloud storage infrastructure add a high cost to each I/O. For example, as disclosed in a white-box study~\cite{zhang2024s}, I/Os in Alibaba EBS traverse a two-hop network (from BlockClient to BlockServer, then to ChunkServer) and a multi-faceted software stack (BlockClient, BlockServer, and Pangu).
Second, the lack of a carefully-coordinated rate-limiting mechanism further exacerbates latency, as we will detail in Section~\ref{sec:model}.


Furthermore, latency gaps are remarkably smaller in random read workloads compared to the other three types (i.e., random write, sequential write, and sequential read).
For ESSD-AM1, the average and P99.9 latency gaps in random read reach only 6.3$\times$ and 8.7$\times$, whereas in the other workloads they climb as high as 22.7$\times$ and 14.7$\times$ (random write), 21.7$\times$ and 14.9$\times$ (sequential write), and 39.3$\times$ and 36.4$\times$ (sequential read).


This discrepancy arises because modern SSDs often use a DRAM-based write buffer to accelerate writes, and prefetching to improve sequential read performance~\cite{li2022fantastic}.
As a result, random/sequential writes and sequential reads incur fewer actual flash accesses, making the cloud-inherent overhead more noticeable.
Random reads, however, benefit less from prefetching and involve more flash operations, thereby reducing the relative impact of network and software latency and resulting in a smaller gap.


These findings indicate that, to avoid severe latency penalties in EBS environments, users should \textit{avoid small and outstanding I/Os on the critical path} (\textbf{Implication\jc}).
Notably, when I/Os are adequately scaled in write-intensive workloads, ESSD-AM1 achieves up to 25\% lower average latency and 3.3$\times$ lower P99.9 latency than the local SSD, as shown in Figure~\ref{exp_lat_high}b.


\subsection{Garbage Collection (GC)} \label{sec:uc_gc}

\vspace{.5em}
\noindent\fbox{\parbox{0.475\textwidth}{
\textbf{Observation\cc:} \emph{The user-visible performance degradation commonly associated with device-side GC is substantially delayed or not observed.}
}}
\vspace{.5em}

Device-side garbage collection (GC) is widely acknowledged as a major source of SSD performance degradation: the FTL periodically relocates valid data to reclaim space, and these extra writes compete with foreground I/Os.
Contrary to these conventional expectations, however, our experiments reveal that the user-visible performance degradation commonly associated with device-side GC is substantially delayed or not observed.

\begin{figure}[t]
    \centering
    \includegraphics[width=0.46\textwidth]{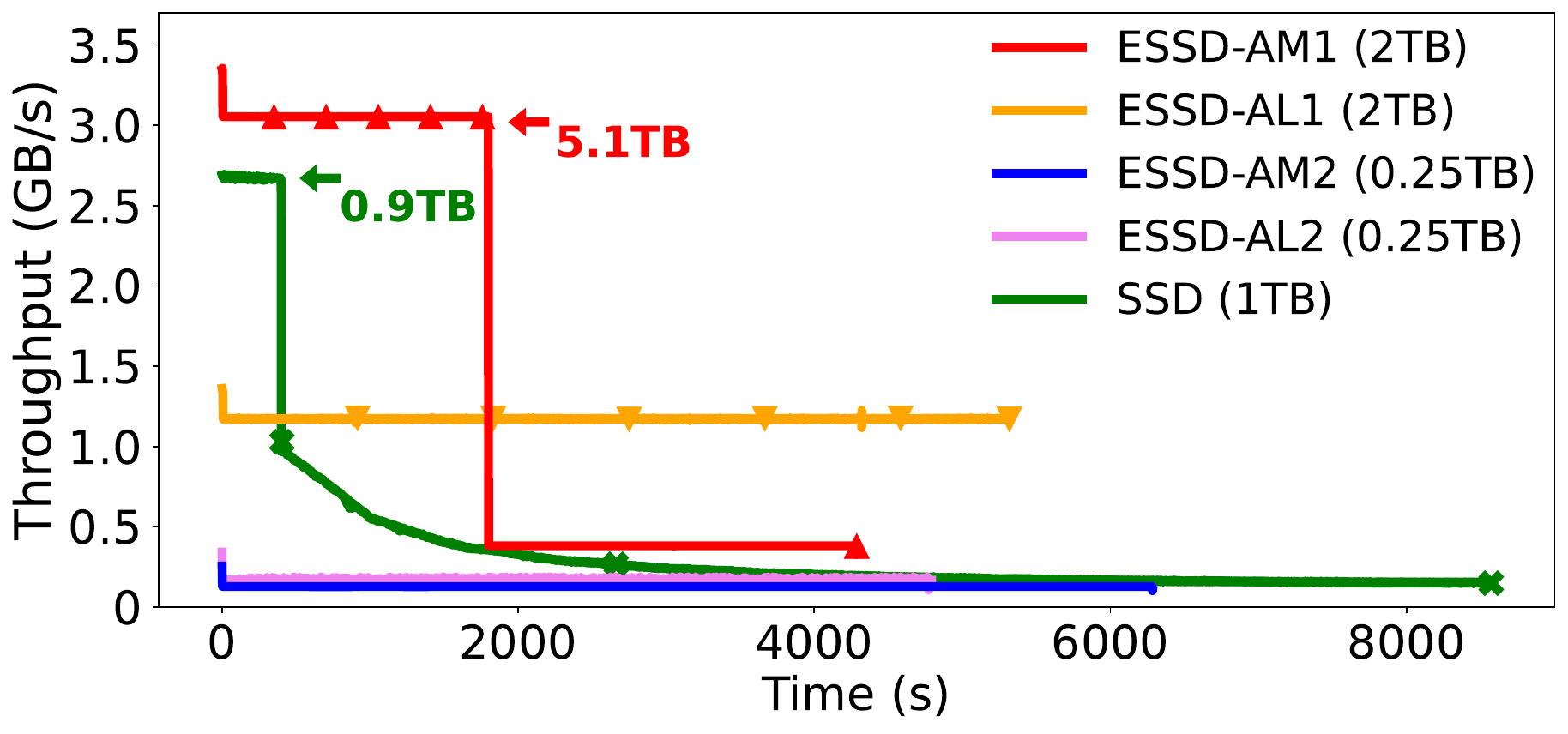}
    \caption{
    \textbf{Runtime throughput of four ESSDs and SSD under a random write workload with four I/O streams until writing 3$\times$ storage capacity.}
    Thus, the write volume applied to each device is 3$\times$ its rated capacity, resulting in 6TB per high-end ESSD, 0.75TB per low-end ESSD, and 3TB for the local SSD.
    The markers indicate the time when the total write volume reaches a multiple of 1TB.
    }
    \label{exp_gc}
\end{figure}

Figure~\ref{exp_gc} presents the runtime throughput of four ESSD instances (i.e., high-end ESSD-AM1 from Amazon AWS, high-end ESSD-AL1 from Alibaba Cloud, low-end ESSD-AM2 from Amazon AWS, and low-end ESSD-AL2 from Alibaba Cloud) along with a local SSD Samsung 970 Pro, under a random write workload with four concurrent I/O streams.
The total amount of data written was set to three times the storage capacity of each device.


As expected, the local SSD exhibits a sharp throughput drop after writing 0.9TB (90\% of its capacity), where performance falls by 63\% from 2.7GB/s to 1.0GB/s.
The throughput continues to decline with additional writes, eventually reaching as low as 149MB/s.
This extended period of low performance highlights the vulnerability of local SSDs to garbage collection under sustained write pressure.


In contrast, ESSDs maintain high throughput over significantly longer durations, even under intense write loads, even though the exact behavior varies by provider.
Most ESSDs, including ESSD-AM2 and both Alibaba Cloud instances, sustain consistent performance until writes reach 3$\times$ their capacity.
Only ESSD-AM1 experiences a sudden throughput reduction after 5.1TB (2.55$\times$ its capacity), eventually stabilizing around 305MB/s.
We attribute this drop to the exhaustion of instance-level EBS burst credits rather than device-side GC~\cite{website:awsm6in}. 
At the measured throughput, writing 5.1TB takes approximately 27 minutes, closely matching the documented 30-minute burst window of \texttt{m6in.xlarge}; the post-drop throughput is also close to its baseline EBS bandwidth~\cite{website:awsm6in}.



Thus, none of the evaluated ESSDs exhibits the progressive throughput degradation observed on the local SSD.
Since the EBS internals remain opaque, our results do not imply that backend GC is absent; rather, they show that its performance impact is not exposed at the user-visible block interface within the tested write volume.


Extensive research focuses on mitigating GC effects in local SSDs through techniques such as bandwidth/space over-provisioning and I/O redirection~\cite{skourtis2014flash,kim2019alleviating,jiang2021fusionraid}.
However, our findings suggest that cloud software designers should \textit{re-evaluate whether existing GC-mitigation techniques used for local SSDs remain necessary on ESSDs} (\textbf{Implication\jc}), due to performance and/or cost trade-offs they have to entail.


\subsection{Access Pattern} \label{sec:uc_acp}

\begin{figure*}[t]
    \centering
    \subfigure[]{
    \includegraphics[width=0.315\textwidth]{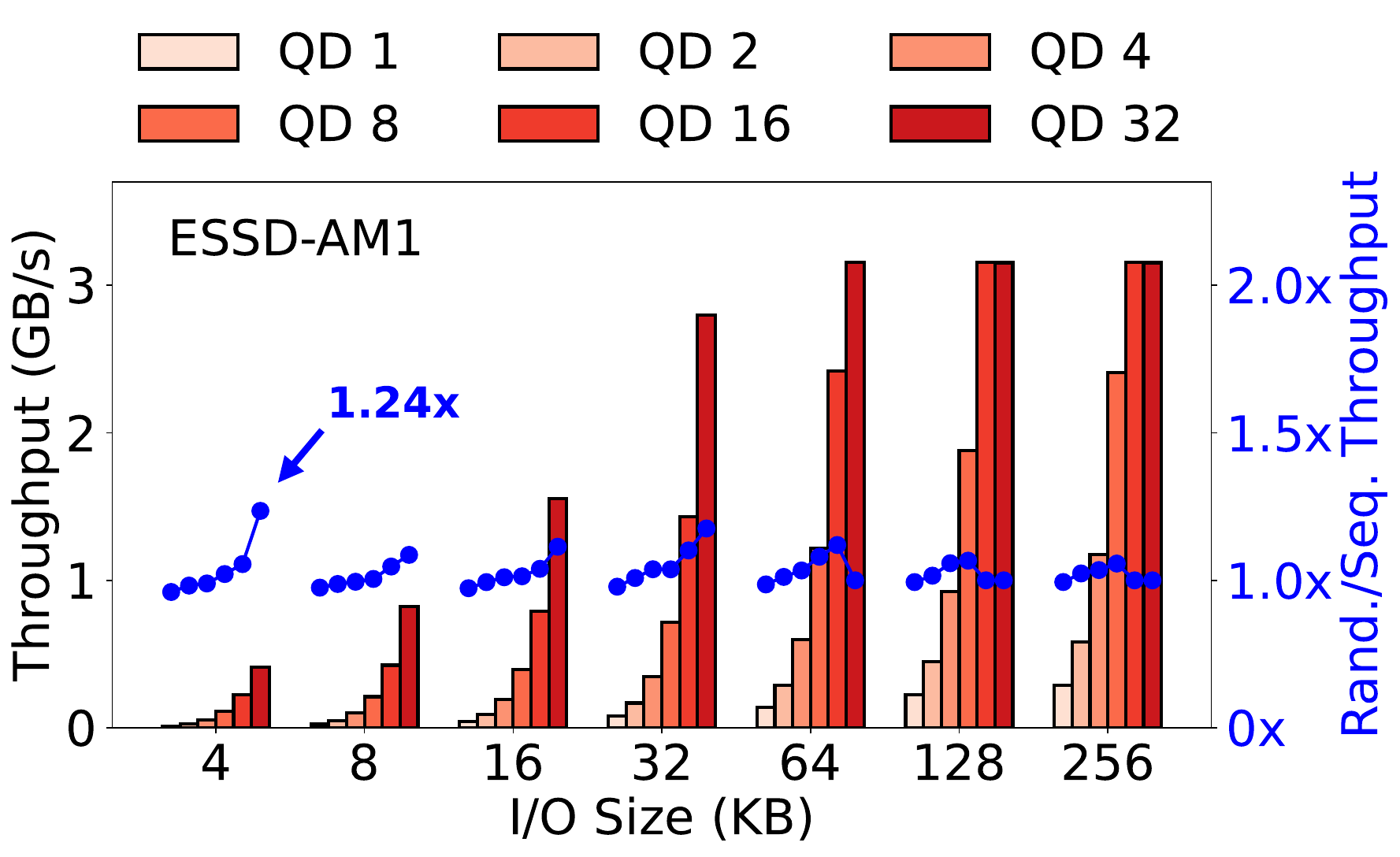}
    }
    \subfigure[]{
    \includegraphics[width=0.315\textwidth]{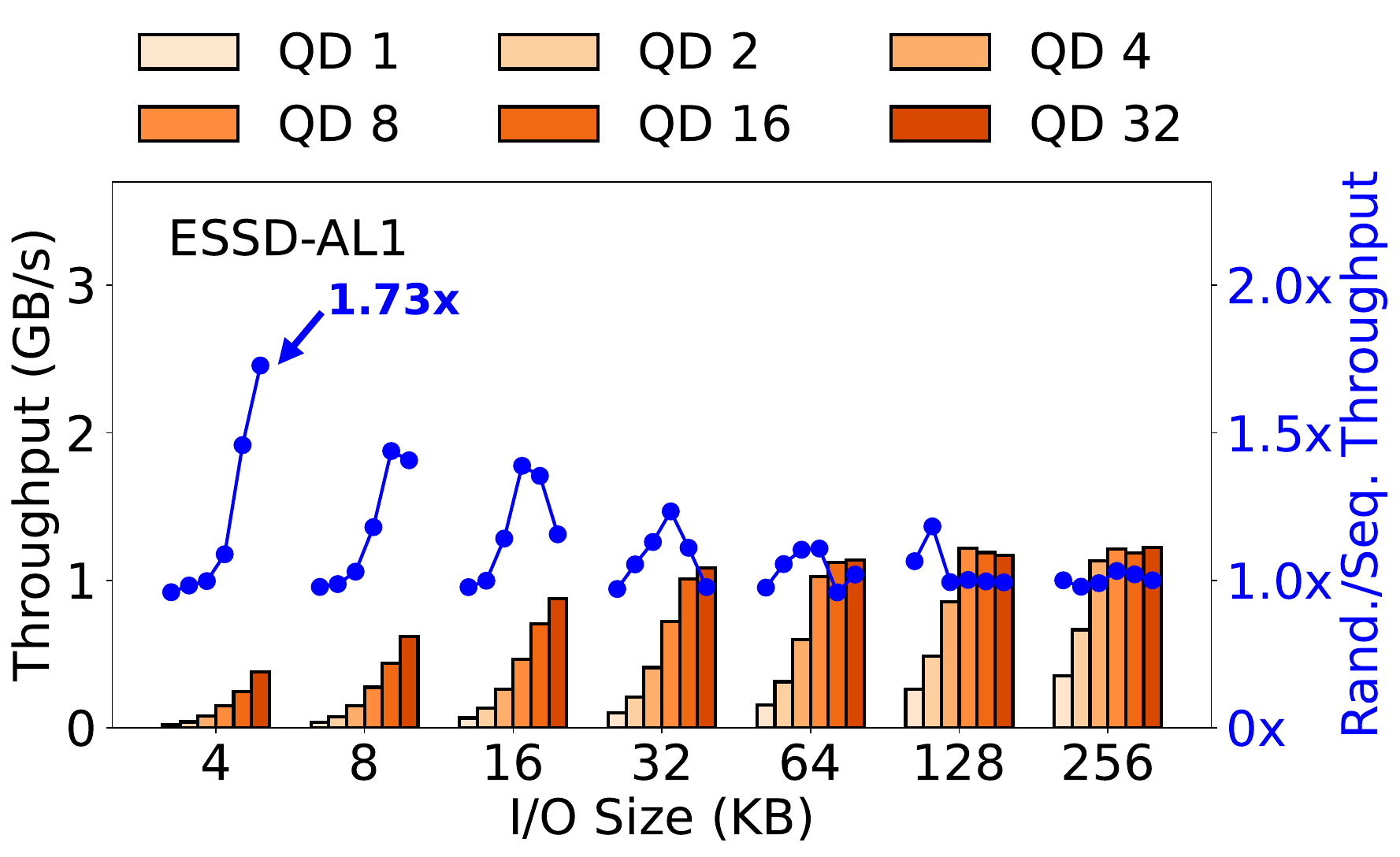}
    }
    \subfigure[]{
    \includegraphics[width=0.315\textwidth]{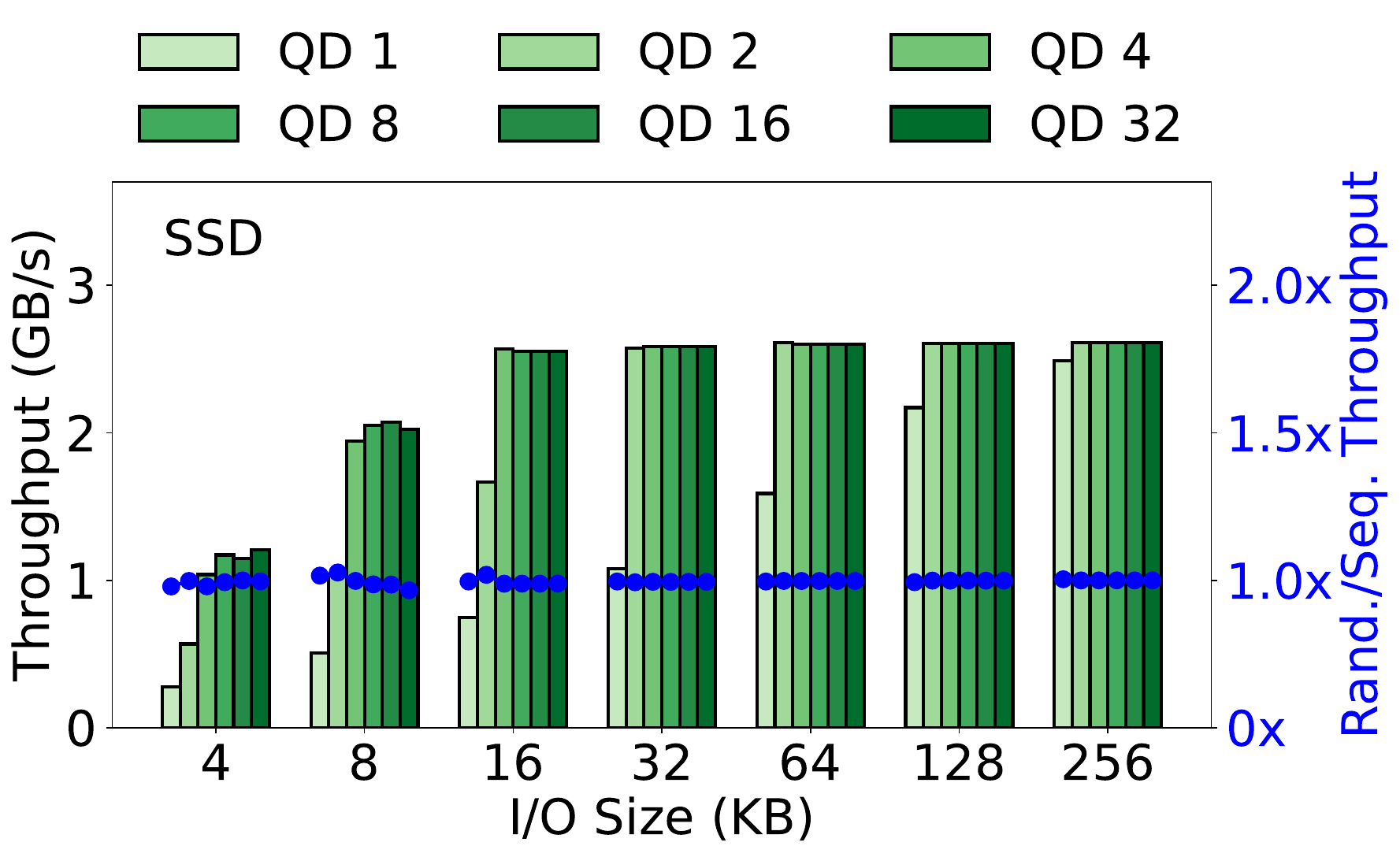}
    }
    \caption{
    \textbf{Throughput of ESSDs and SSD under random write workloads and throughput gain of ESSDs and SSD in random writes over sequential writes.}
    The I/O size varies from 4KB to 256KB, and the I/O queue depth (QD) varies from 1 to 32.
    Note that for clarity, different subfigures have different y-axis scales.
    }
    \label{exp_bw}
\end{figure*}

\vspace{.5em}
\noindent\fbox{\parbox{0.475\textwidth}{
\textbf{Observation\cc:} \emph{In high-end ESSDs, the throughput of random writes outperforms that of sequential writes, reaching a maximum of 1.24$\times$ and 1.73$\times$ in the two ESSDs, respectively.}
}}
\vspace{.5em}

Random writes have traditionally been viewed as detrimental to SSD performance, primarily because they increase the interleaving of valid and invalid data within flash blocks, thereby raising the overhead of relocating valid data during garbage collection.
Surprisingly, however, our experiments reveal that high-end ESSDs often achieve higher throughput under random writes compared to sequential writes, but this phenomenon is absent in low-end ESSDs.


As illustrated in Figure~\ref{exp_bw}, we examine the throughput advantage of ESSDs and a local SSD in random writes relative to sequential writes (represented by blue lines).
For clarity, we refer to this measure as \textit{throughput gain} throughout this subsection.
The tests span I/O sizes from 4KB to 256KB and queue depths from 1 to 32.
Note that the y-axis scales differ across subfigures for better visual comparison.


A key observation is that two high-end ESSDs (i.e., ESSD-AM1 and ESSD-AL1) exhibit throughput gains of up to 1.24$\times$ and 1.73$\times$, respectively.
In contrast, the local SSD shows negligible performance differences between random and sequential writes (when GC does not occur).
This divergence may stem from the distributed storage architecture of ESSDs. Because data is striped and replicated across multiple nodes and physical SSDs (Section~\ref{sec:bg_essd}), random writes can exploit aggregated bandwidth from multiple physical SSDs, thereby improving user-perceived throughput.


Meanwhile, throughput gains for the two low-end ESSDs (i.e., ESSD-AM2 and ESSD-AL2) remain near 1, indicating no notable performance benefit from random writes.
This is likely due to their constrained performance budgets: specifically, ESSD-AM2 offers only 4.2\% of ESSD-AM1's maximum bandwidth and 5\% of its IOPS, while ESSD-AL2 provides 16.4\% of ESSD-AL1's maximum bandwidth and 4.8\% of its IOPS (see Table~\ref{essd_config}).
Such limitations cause them to saturate the budget quickly, leaving little room for further gains.


We also observe provider-specific variations in throughput gains.
ESSD-AM1 from Amazon AWS shows moderate improvement, mainly under high queue depths and small-to-medium I/O sizes, for example, a 23.5\% gain at 4KB and queue depth 32.
ESSD-AL1 from Alibaba Cloud, however, demonstrates more substantial gains, reaching up to 72.8\%, 43.8\%, 38.8\%, and 23.4\% for I/O sizes of 4KB, 8KB, 16KB, and 32KB, respectively.


Conventionally, transforming random writes into sequential writes through methods such as log-structuring~\cite{lee2015f2fs} or copy-on-write~\cite{rodeh2013btrfs} has been advocated to reduce GC overhead and improve SSD performance.
However, given that ESSDs show considerable tolerance to GC and may even benefit from random writes, it is worth \textit{rethinking whether it is still worthwhile to convert random writes in random-write-based software into sequential writes and if it is beneficial to proactively trigger random writes in sequential-write-based software} (\textbf{Implication\jc}).


\subsection{Performance Budget} \label{sec:uc_mix}

\vspace{.5em}
\noindent\fbox{\parbox{0.475\textwidth}{
\textbf{Observation\cc\ (validating the written contract):} \emph{The maximum bandwidth and IOPS guaranteed by the provider are experimentally validated to be deterministic and no longer sensitive to the access pattern.}}
}
\vspace{.5em}

\begin{figure*}[t]
    \centering
    \subfigure[Throughput]{
    \includegraphics[width=0.48\textwidth]{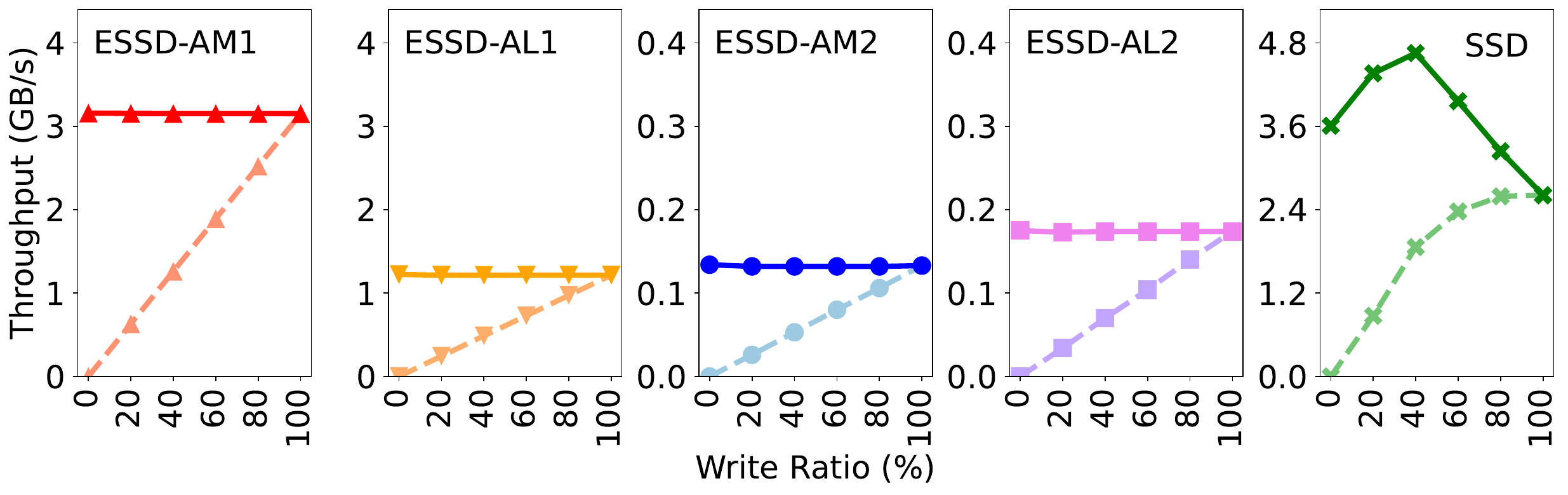}
    }
    \subfigure[IOPS]{
    \includegraphics[width=0.48\textwidth]{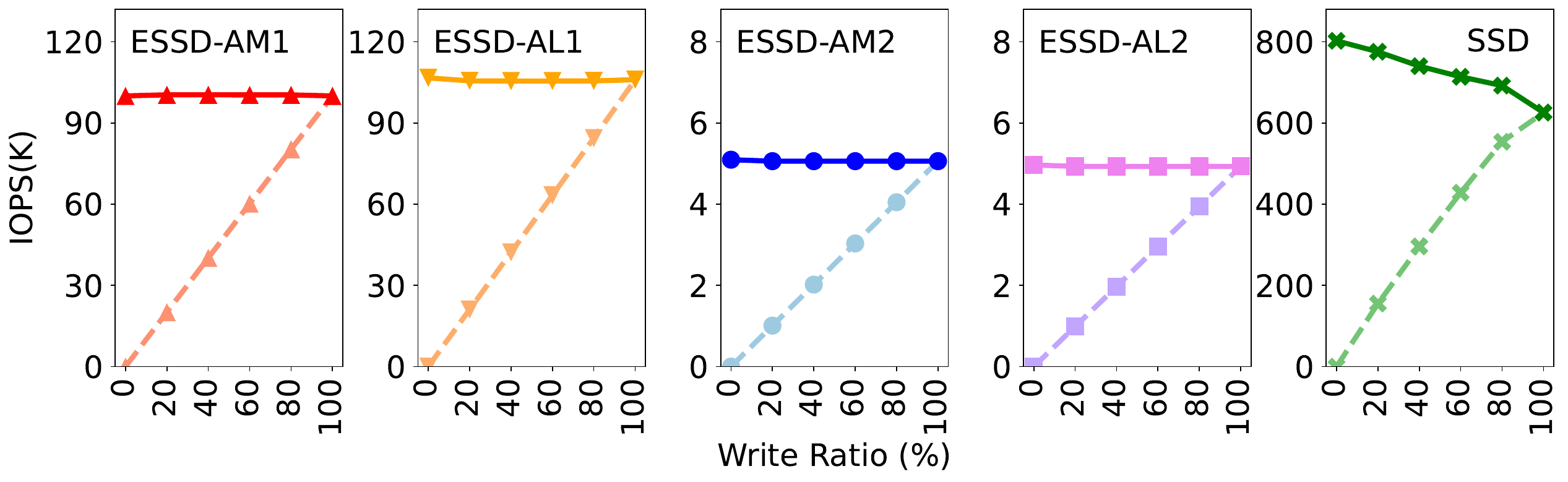}
    }
    \caption{
    \textbf{Throughput and IOPS of four ESSDs and SSD under mixed read/write workloads with different write ratios.}
    The solid and dashed lines represent the total (i.e., read and write) throughput or IOPS and the write throughput or IOPS, respectively.
    Note that for clarity, different subfigures have different y-axis scales.
    }
    \label{exp_mix}
\end{figure*}

Local SSD peak throughput varies considerably with access pattern due to asymmetric flash operation latencies (e.g., 3.5GB/s read vs.\ 2.7GB/s write in Samsung 970 Pro, see Table~\ref{essd_config}).
In contrast, subject to the active performance limits, the bandwidth and IOPS budgets observed for ESSDs remain consistent across access patterns.

Figure~\ref{exp_mix} displays the throughput and IOPS of four ESSDs and one local SSD under mixed read/write workloads with write ratios ranging from 0\% (pure random read) to 100\% (pure random write).
The throughput and IOPS experiments use 128KB and 4KB I/O sizes, respectively.
Note that the y-axis scales differ across subfigures for clear visualization.
A clear trend emerges: for all four ESSDs, both throughput and IOPS remain stable and closely align with the values guaranteed by the provider, regardless of write ratio.
Conversely, the local SSD exhibits non-deterministic performance, with throughput fluctuating between 2.6GB/s and 4.6GB/s and IOPS varying from 620K to 800K.
These findings reveal a fundamental difference: unlike local SSDs that require complex performance tuning, ESSDs offer predictable, stable throughput boundaries, which yields two practical implications.

First, since performance budgets directly influence cost (as noted in Section~\ref{sec:bg_essd}), cloud software should \textit{smooth the read/write I/Os to be evenly distributed across the timeline and below the guaranteed performance budget} (\textbf{Implication\jc}).
This is especially critical for burst-heavy applications where load variability complicates capacity planning.

Second, cloud software should also \textit{re-examine I/O reduction techniques (e.g., compression and deduplication) that were historically considered to impair performance} (\textbf{Implication\jc}).
For ESSDs, where network and software processing latencies dominate, the additional compute cost of these methods may no longer be the limiting factor.
Thus, applying I/O reduction can not only lower costs by reducing the required performance budgets but also improve the overall performance.

\section{The I/O Rate-Limiting Model for Cloud Software using ESSDs} \label{sec:model}

In this section, we investigate how to set up the I/O rate-limiting model for cloud software that employs ESSDs, aiming to achieve low latency without unanticipated spikes.
We begin by introducing the current state-of-the-art I/O rate-limiting model and discussing its identified limitations in Section~\ref{sec:model_bg}.
Subsequently, we present our refined approach to I/O rate-limiting in Section~\ref{sec:model_ours}.


\subsection{The State-of-the-Art Work and Its Limitations} \label{sec:model_bg}

Zhou et al. first emphasize the importance of aligning I/O rates to ESSDs with the internal rate-limiting mechanisms of EBS, indicating that surpassing the provider-defined IOPS budget per second can lead to significant latency increases~\cite{zhou2023calcspar}. 
Their IOPS stabilizer uses the \textit{token bucket algorithm}~\cite{website:tokenbucket} to restrict the IOPS issued to EBS, with tokens refreshed every second according to the provisioned IOPS budget.



The token bucket algorithm is a well-established technique for rate regulation and traffic shaping.
The algorithm uses a bucket that holds tokens, each authorizing the processing of one unit of work, such as an I/O request in this context.
Tokens are refilled at a consistent rate, and incoming tasks must acquire a token to be executed.
If insufficient tokens are available, tasks are delayed until enough tokens accumulate.
This mechanism helps maintain system stability while ensuring controlled performance.

In practical implementations, the token bucket algorithm often incorporates support for multiple priority queues, allowing differentiated token distribution according to task criticality.
Under this extended model, several queues, each associated with a priority, compete for tokens from a common bucket.
A typical realization employs weighted fair queuing (WFQ)~\cite{website:wfq}, which allocates more tokens to higher-priority queues to ensure timely processing of urgent tasks, while still guaranteeing a minimal share for lower-priority queues to prevent starvation.
By incorporating task priority, the token bucket method becomes more adaptive to real-world system demands and enhances overall operational efficiency.



\begin{figure}[t]
    \centering
    \subfigure[Bandwidth]{
    \includegraphics[width=0.45\textwidth]{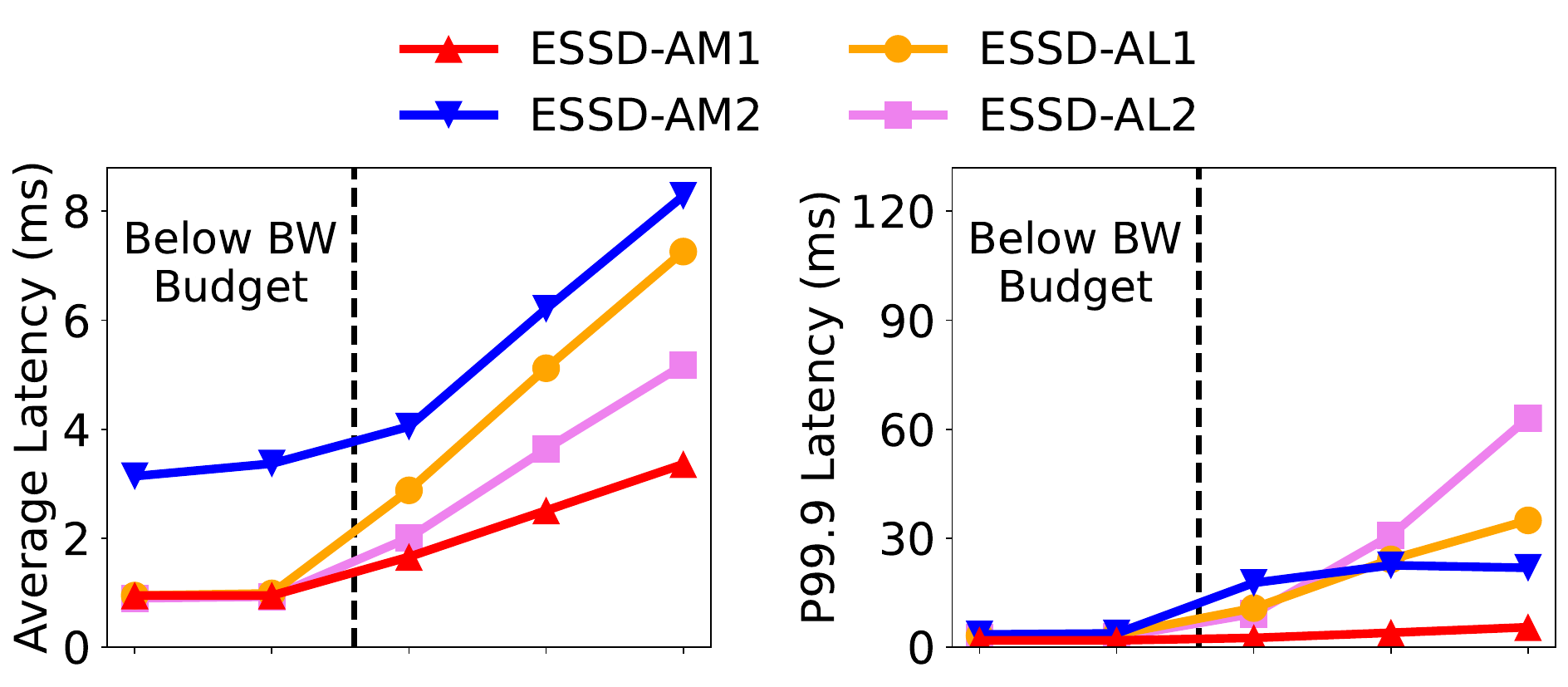}
    }
    \subfigure[IOPS]{
    \includegraphics[width=0.45\textwidth]{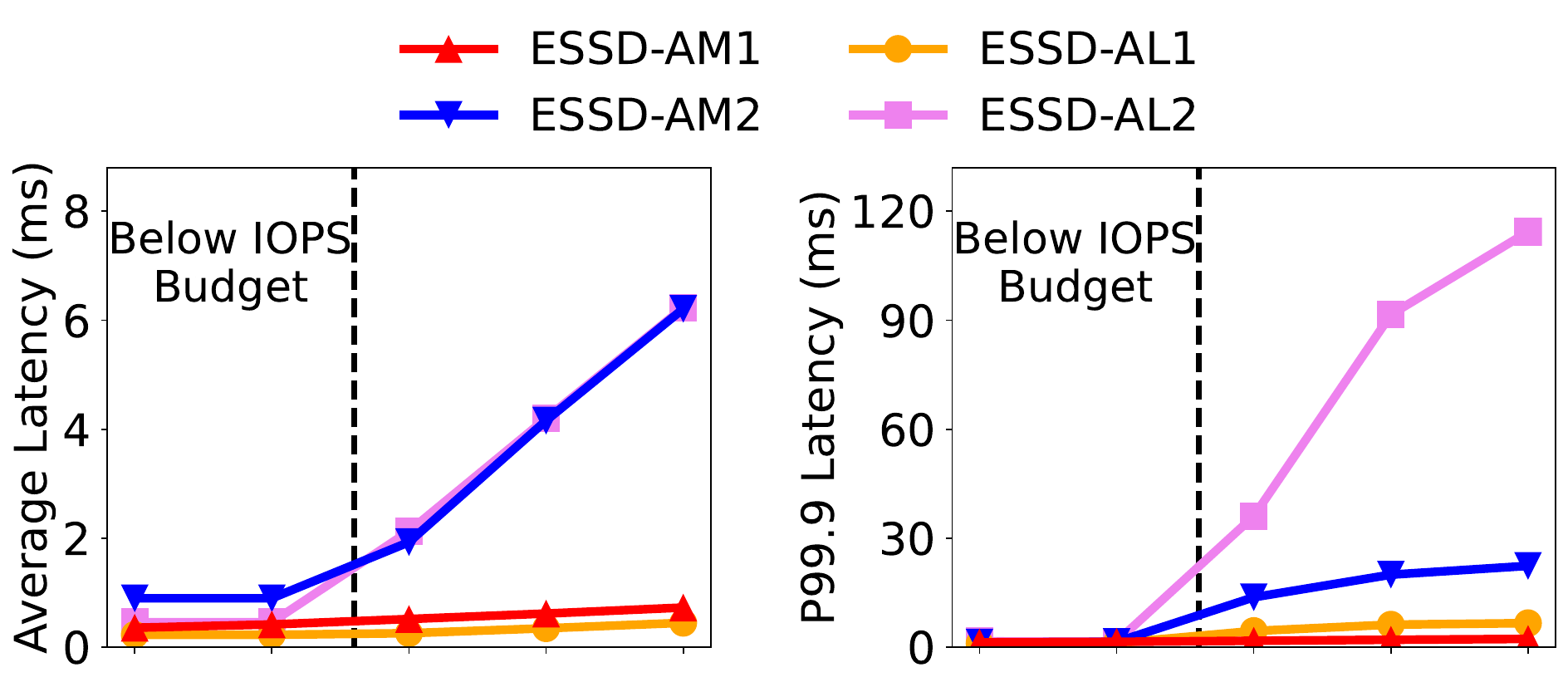}
    }
    \caption{
    \textbf{Average latency and P99.9 latency of four ESSDs under write workloads with different bandwidth or IOPS intensities.}
    Specifically, the groups to the left of the black dashed line indicate that their bandwidth or IOPS requirements are below that guaranteed by the provider, and the groups to the right indicate that they are above.
    The closer to the left, the group with a lower bandwidth/IOPS requirement.
    Note that for clarity, different subfigures have different y-axis scales.
    }
    \label{limit1}
\end{figure}

Although representing an important step forward for cloud software development, our analysis identifies three major limitations in their I/O rate-limiting model.
The first limitation relates to \textit{which type of rate should be limited} (Section~\ref{sec:model_limit1}).
Specifically, we observe that restricting only IOPS is inadequate, as surpassing the allocated bandwidth budget can also lead to substantially increased latency.
The second limitation involves \textit{the method used to enforce rate limits} (Section~\ref{sec:model_limit2}).
In particular, our experimental results indicate that the token refilled interval for rate-limiting varies by provider, and overlooking this factor can prevent cloud software from achieving optimal latency.
The third limitation is that \textit{a latency spike is observed due to non-instantaneous synchronization between cloud-software-side rate-limiting and EBS-side rate-limiting} (Section~\ref{sec:model_limit3}).
We elaborate on each of these issues in the subsequent sections.


\subsubsection{Limitation \#1: Bandwidth Shortage Also Matters} \label{sec:model_limit1}

Figure~\ref{limit1} illustrates the average and 99.9th percentile (P99.9) latency of four ESSDs under write workloads with varying bandwidth and IOPS intensities (specifically, whether these intensities exceed the allocated budgets).
To generate these workloads, we also use the FIO benchmarking tool, adjusting bandwidth and IOPS levels by modifying (1) the number of I/O threads and (2) the number of I/Os issued per thread per token refilled interval (via \texttt{thinktime} and \texttt{thinktime\_blocks} parameters).
The I/O sizes are set to 256KB and 4KB for bandwidth-related tests and IOPS-related experiments, respectively.
The token refilled interval is configured to 1 second for Amazon AWS ESSDs and 20 milliseconds for Alibaba Cloud ESSDs, as further elaborated in Section~\ref{sec:model_limit3}.

The results indicate that latency metrics deteriorate markedly when workload demands exceed the provider-guaranteed bandwidth or IOPS budgets.
In IOPS-related tests (Figure~\ref{limit1}b), the two low-end ESSDs (i.e., ESSD-AM2 and ESSD-AL2) exhibit average latency increases of 6.9$\times$ and 13.5$\times$, and P99.9 latency increases of 15.1$\times$ and 73.3$\times$ under IOPS oversubscription.
Even high-end ESSDs (i.e., ESSD-AM1 and ESSD-AL1) are affected, with average latency rising by 73.8\% and 95.7\%, and P99.9 latency increasing by 1.5$\times$ and 5.7$\times$.
These findings underscore the necessity of controlling I/O counts per interval to achieve consistent latency, corroborating earlier observations in~\cite{zhou2023calcspar}.

But notably, bandwidth constraints also have a substantial impact.
As depicted in Figure~\ref{limit1}a, low-end ESSDs (i.e., ESSD-AM2 and ESSD-AL2) experience average latency increases of 2.5$\times$ and 5.6$\times$, and P99.9 latency increases of 5.7$\times$ and 22.4$\times$ when bandwidth budgets are exceeded.
High-end devices (i.e., ESSD-AM1 and ESSD-AL1) are similarly impaired, showing average latency growth of 3.5$\times$ and 7.3$\times$, and P99.9 latency elevations of 2.9$\times$ and 9.8$\times$.
\textit{These outcomes demonstrate that surpassing bandwidth limits considerably degrades performance, indicating that cloud software must regulate not only IOPS but also bandwidth.
}

\begin{figure}[t]
    \centering
    \subfigure[Bandwidth]{
    \includegraphics[width=0.45\textwidth]{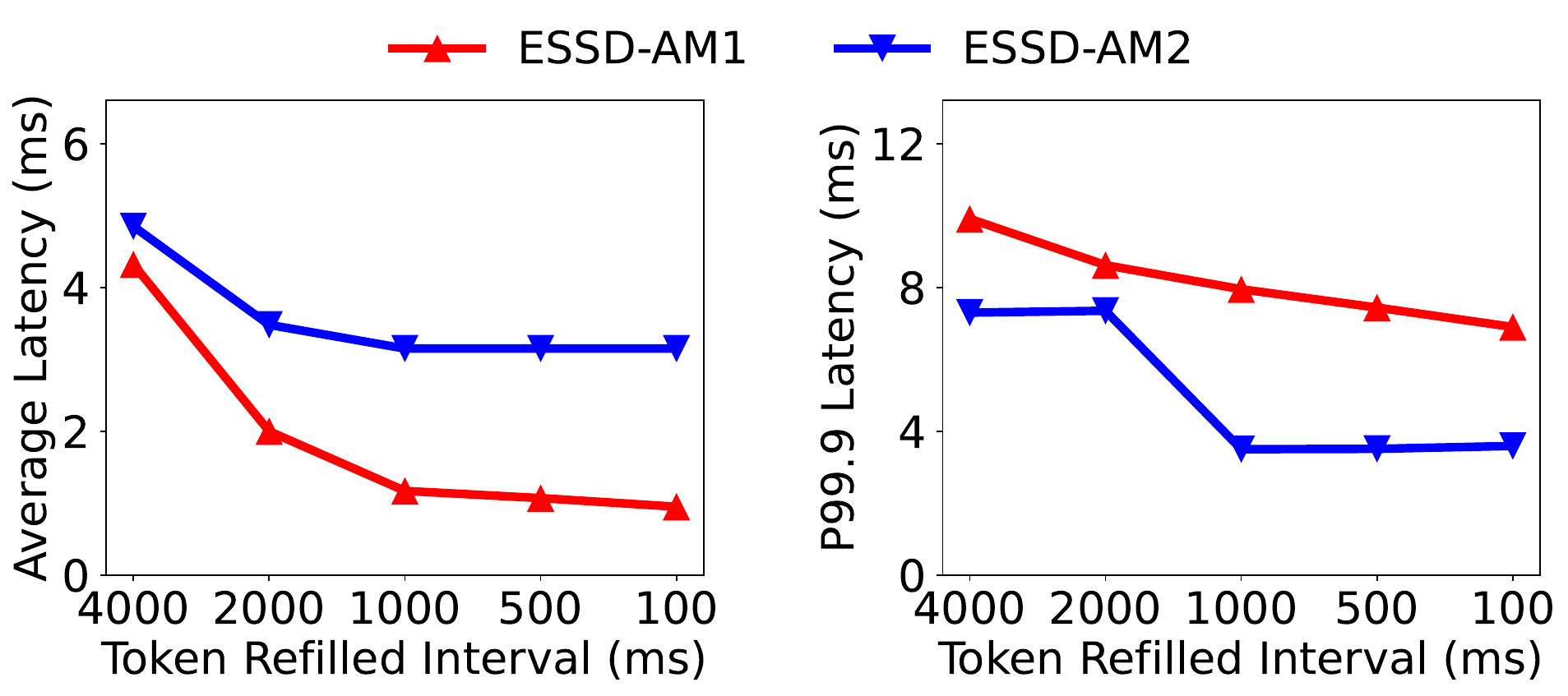}
    }
    \subfigure[IOPS]{
    \includegraphics[width=0.45\textwidth]{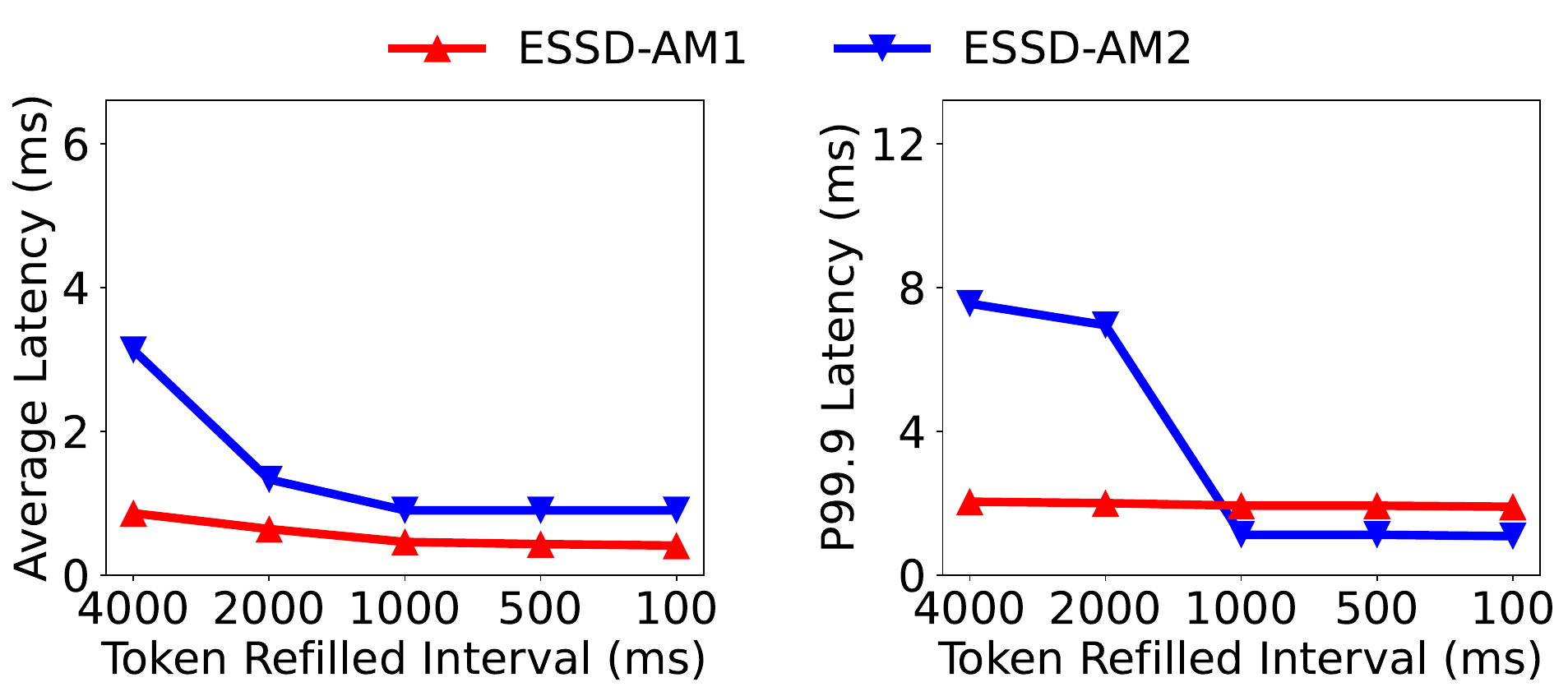}
    }
    \caption{
    \textbf{Average latency and P99.9 latency of Amazon AWS ESSDs under write workloads with different token refilled intervals.}
    The bandwidth or IOPS intensities of the workloads are close but slightly lower than the budget of each type of ESSD.
    Note that for clarity, different subfigures have different y-axis scales.
    }
    \label{limit2_aws}
\end{figure}

\subsubsection{Limitation \#2: Token Refilled Interval is not Static} \label{sec:model_limit2}

\begin{figure}[t]
    \centering
    \subfigure[Bandwidth]{
    \includegraphics[width=0.48\textwidth]{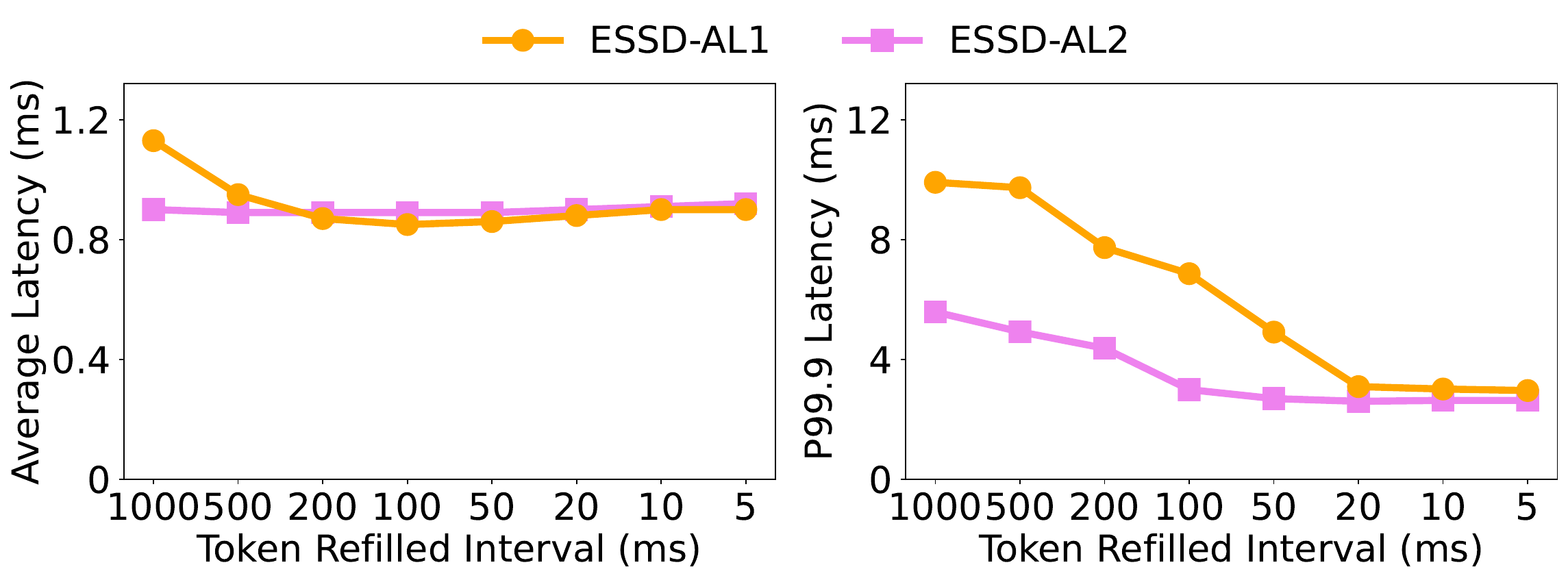}
    }
    \subfigure[IOPS]{
    \includegraphics[width=0.48\textwidth]{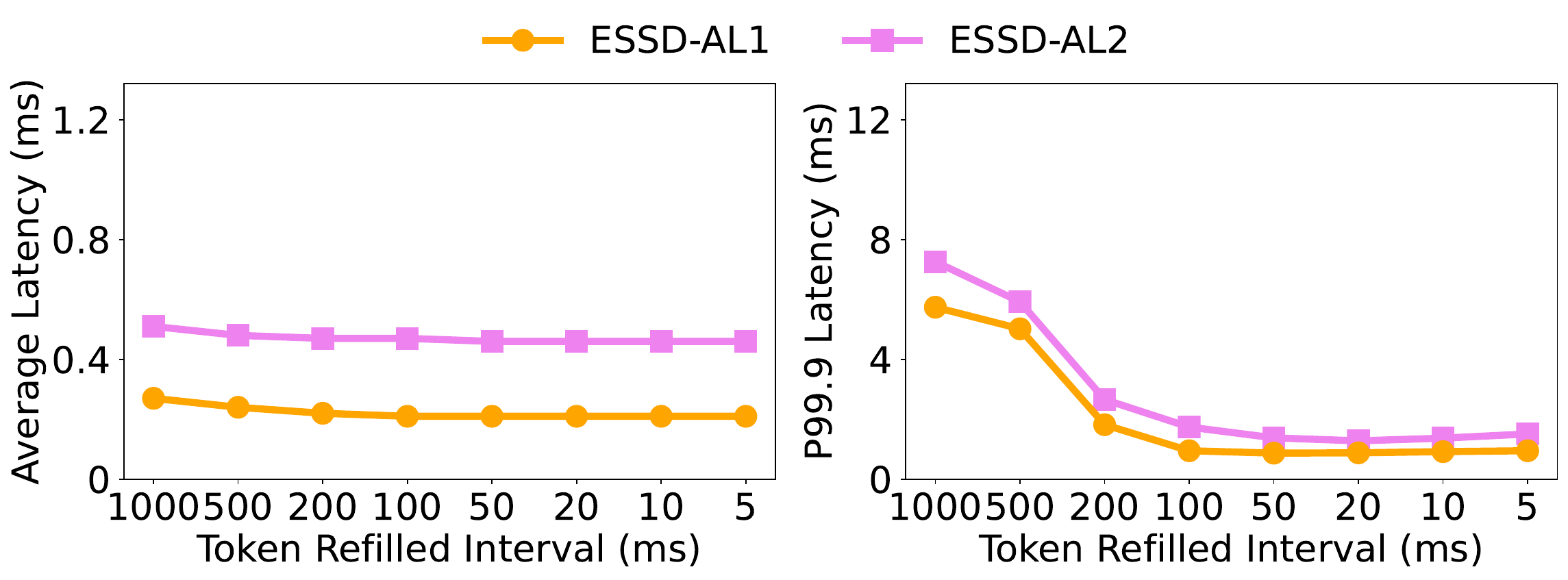}
    }
    \caption{
    \textbf{Average latency and P99.9 latency of Alibaba Cloud ESSDs under write workloads with different token refilled intervals.}
    The bandwidth or IOPS intensities of the workloads are close but slightly lower than the budget of each type of ESSD.
    Note that for clarity, different subfigures have different y-axis scales.
    }
    \label{limit2_ali}
\end{figure}

Figures~\ref{limit2_aws} and~\ref{limit2_ali} show the average latency and P99.9 latency of ESSDs with different token refilled intervals from Amazon AWS and Alibaba Cloud, respectively.
Specifically, the token refilled intervals are 4000ms, 2000ms, 1000ms, 500ms, and 100ms. 
We still use the FIO benchmark tool to generate workloads, and the bandwidth or IOPS intensities are controlled to be close but slightly lower than the budget of each type of ESSD.
For bandwidth- and IOPS-related experiments, the I/O size is still set to 256KB and 4KB, respectively.

Results in Figure~\ref{limit2_aws} indicate that latency stabilizes for Amazon AWS ESSDs once the token refilled interval is set to 1000ms (1 second) or shorter, a trend consistent across both bandwidth and IOPS experiments.
Particularly, for bandwidth-related tests (Figure~\ref{limit2_aws}a), reducing the interval from 4000ms to 1000ms results in a 75.2\% decrease in average latency and a 21.9\% reduction in P99.9 latency for ESSD-AM1, while ESSD-AM2 showed improvements of 35.1\% in average latency and 51.9\% in P99.9 latency.
Similarly, in IOPS-related experiments (Figure~\ref{limit2_aws}b), ESSD-AM1 exhibits a 50.0\% drop in average latency and a 5.4\% decrease in P99.9 latency, whereas ESSD-AM2 sees more substantial gains of 71.2\% and 85.0\%, respectively.


These results can verify that the token refilled interval in Amazon AWS ESSDs is highly likely to be 1 second.
As a result, configuring a one-second token refilled interval in the cloud software I/O rate-limiting model can yield good latency in Amazon AWS ESSDs.

However, we found that the one-second token refilled interval fails to achieve the best performance on Alibaba Cloud ESSDs.
Figure~\ref{limit2_ali} shows the latency of two Alibaba Cloud ESSDs at token refilled intervals from 1000ms to as small as 5ms.
We can see that both the average latency and P99.9 latency of the two ESSDs can be further improved if the token refilled interval is configured less than 1000ms.
Specifically, regarding the high-end ESSD-AL1, the average latency in bandwidth-related experiments (Figure~\ref{limit2_ali}a) and both the average latency and P99.9 latency in IOPS-related experiments (Figure~\ref{limit2_ali}b) become stable when the token refilled interval is decreased to 100ms.
Similarly, for the low-end ESSD-AL2, the average or P99.9 latency in bandwidth- or IOPS-related experiments gets steady at the token refilled interval of 100ms.

These results suggest that the token refilled interval within Alibaba Cloud EBS is likely to be 100ms, so configuring a one-second interval on the cloud software side would lead to unnecessary performance loss.
Taking ESSD-AL1 as an example, the average latency and P99.9 latency are improved by up to 44.5\% and 5.0$\times$, respectively.

\textit{The above results can convey that the token refilled interval for rate-limiting is actually provider-specific.}
That being said, the cloud software I/O rate-limiting model should determine the exact token refilled interval with the underlying provider in mind and profiling experiments beforehand.
The experimental methodology in this section could be useful for reference.

\subsubsection{Limitation \#3: Non-Instantaneous Synchronization of Cloud-Software-Side Rate-Limiting with EBS-Side Rate-Limiting} \label{sec:model_limit3}

Meanwhile, we can also notice a discrepancy in Figure~\ref{limit2_ali}a, i.e., the P99.9 latency of ESSD-AL1 in bandwidth-related experiments becomes stable only when the token refilled interval is decreased to 20ms.
The primary cause of this smaller token refilled interval could be the non-instantaneous synchronization of cloud-software-side rate-limiting with EBS-side rate-limiting, where the time differences of rate-limiting at both ends could unexpectedly block some I/Os and increase the tail latency.
This problem is more challenging (and may more likely appear) in bandwidth-related experiments because bandwidth experiments employ 256KB I/Os, which may incur higher processing overhead (compared to 4KB I/Os in IOPS experiments) and are thus more likely to increase the time difference between both ends.
In addition, high-end ESSDs require supporting a bandwidth of a few GB/s (e.g., 1.1GB/s in ESSD-AL1), which can also exacerbate this problem because it is intrinsically hard for the provider to manage a massive number of I/Os while not affecting tail latency.

To conclude, there exists a non-instantaneous synchronization between cloud-software-side rate-limiting and EBS-side rate-limiting, which may adversely impact performance in workloads with large-sized I/Os and high-end ESSDs.
Meanwhile, we found that setting up a finer-grained (or smaller) token refilled interval on the cloud software side is a feasible solution to address this problem without degrading the latency.
For example, when the token refilled interval is configured to 20ms (versus 100ms), the P99.9 latency of ESSD-AL1 in bandwidth-related experiments is decreased by 59.8\% (see Figure~\ref{limit2_ali}a).

\subsection{Our Solution} \label{sec:model_ours}

Based on the above experiments and analysis, in this section, we present a refined I/O rate-limiting model for cloud software using ESSDs to obtain satisfactory latency.
Our solution introduces two architectural changes beyond Calcspar's IOPS-only, one-second-refill rate-limiting model~\cite{zhou2023calcspar}.

\begin{algorithm}
\caption{Pseudo-code of bandwidth-IOPS dual limiting} \label{pre}
\begin{algorithmic}[1]
\Function{ProcessRequest}{req}
    \State // A request can only be processed if enough tokens are acquired from both bandwidth and IOPS rate limiters

    \State // Step 1: Acquire tokens for bandwidth
    \State $\text{remainingBytes} \gets \text{req.sizeInBytes}$
    \While{$\text{remainingBytes} > 0$}
        \State $\text{acquiredBytes} \gets \text{bandwidthRateLimiter.acquire(remainingBytes)}$
        \If{$\text{acquiredBytes} > 0$}
            \State $\text{remainingBytes} \gets \text{remainingBytes} - \text{acquiredBytes}$
        \EndIf
    \EndWhile
    \State
    \State // Step 2: Acquire token for IOPS
    \While{$\text{!iopsRateLimiter.acquire(1)}$}
    \EndWhile
    
    \State
    \State // Continue to process the request
    \State // ...
\EndFunction
\end{algorithmic}
\end{algorithm}

The first is \textit{bandwidth-IOPS dual limiting}, which indicates that both the bandwidth and IOPS (rather than only IOPS) should be carefully regulated, in coordination with the rate-limiting model within EBS.
Algorithm~\ref{pre} shows the pseudo-code. Cloud software maintains two rate limiters, one for bytes and one for I/O count. An incoming request must acquire tokens from both limiters before proceeding; if either limiter lacks tokens, the request is deferred.

The second is \textit{fine-grained token refilling}, which has two implications.
On the one hand, the token refilled interval should be meticulously profiled (rather than a fixed one-second) and may vary between different providers.
We show how to figure out this information through profiling experiments in Section~\ref{sec:model_limit2}.
On the other hand, the token refilled interval should be finer than the profiled value, so as to avoid unexpected latency spikes due to non-instantaneous synchronization between cloud-software-side rate-limiting and EBS-side rate-limiting.

\section{Case Study on RocksDB} \label{sec:cs}

To instantiate the contract and the refined rate-limiting model, we conduct an application-level case study based on RocksDB~\cite{website:rocksdb}, a popular LSM-tree-based key-value store.
In Section~\ref{sec:cs_setup}, we go through the general experimental setups shared across the case study.
Then, from Section~\ref{sec:cs_cache} to Section~\ref{sec:cs_compress}, we explore how the cloud database using RocksDB as the storage engine can benefit from our findings.
We conclude the case study in Section~\ref{sec:cs_summary}.

\subsection{Experimental Setups} \label{sec:cs_setup}

We evaluate RocksDB on ESSD-AL2 from Alibaba Cloud (see Table~\ref{essd_config}) using the popular YCSB benchmark tool~\cite{cooper2010benchmarking}.
For each experiment, we first preload 10M key-value pairs (each for a 16-byte key and a 1KB value) into the store and then employ 20 threads to issue 5M requests in total.
We configure a 1GB block cache, which can hold approximately 10\% of the dataset, and use direct I/Os to bypass the page cache effects.
We implement our refined I/O rate-limiting model into RocksDB, namely \texttt{RocksDB-E}, with the bandwidth and IOPS budgets configured at 162MB/s and 4.5K IOPS, both representing 90\% of the ESSD performance budgets (see Table~\ref{essd_config}).
The token refilled interval is set to 20ms, according to the conclusion in Section~\ref{sec:model_limit3}.
As with the default configuration in RocksDB~\cite{website:ratelimiter}, the priority weighting ratio between foreground I/Os (i.e., user reads) and background I/Os (i.e., flushes and compactions) is 10:1, and for background I/Os, that between flushes and compactions is also 10:1.

\subsection{Cache Miss Minimization (Implication \#1)} \label{sec:cs_cache}

Implication~\#1 indicates that the per-I/O cost on ESSDs is substantially higher than on local SSDs, particularly for small outstanding I/Os.
For RocksDB, the most latency-sensitive I/O path is the synchronous user reads~\cite{wang2026making}, where each cache miss triggers small read I/Os to the ESSD with an unavoidable latency penalty of hundreds or thousands of microseconds.
Consequently, minimizing the cache miss rate is no longer merely a performance optimization but becomes a key component for meeting user SLAs.

To this end, techniques that can minimize cache misses are highly recommended for use with ESSDs.
Such techniques typically include allocating larger cache space, pinning index and filter blocks in cache, and even prepopulating data blocks into the block cache after compactions (to mitigate the cache invalidation problem~\cite{yang2020leaper}).
In short, trading memory for cache hits is the first line of defense to secure read SLAs on ESSDs.

\subsection{Holistic I/O Regulation (Implication \#1, Implication \#4, and Rate-Limiting Model)} \label{sec:cs_model}

Implications \#1 and \#4, and the study on the rate-limiting model collectively pinpoint the necessity of a holistic I/O regulation.
This is particularly important for RocksDB, which has bursty I/O traffic due to bandwidth-intensive compactions in the background that can easily threaten SLAs of cloud users.
To demonstrate the effectiveness of our proposed rate-limiting model, we compare \texttt{RocksDB-E} with three baselines: (1) vanilla RocksDB without rate-limiting, i.e., \texttt{RocksDB}, (2) RocksDB using Calcspar's one-second IOPS stabilizer, i.e., \texttt{RocksDB-C}~\cite{zhou2023calcspar}, and (3) RocksDB-E with one-second token-refill intervals, i.e., \texttt{RocksDB-E(1s)}, for breakdown analysis.

\begin{figure}[t]
    \centering
    \includegraphics[width=0.48\textwidth]{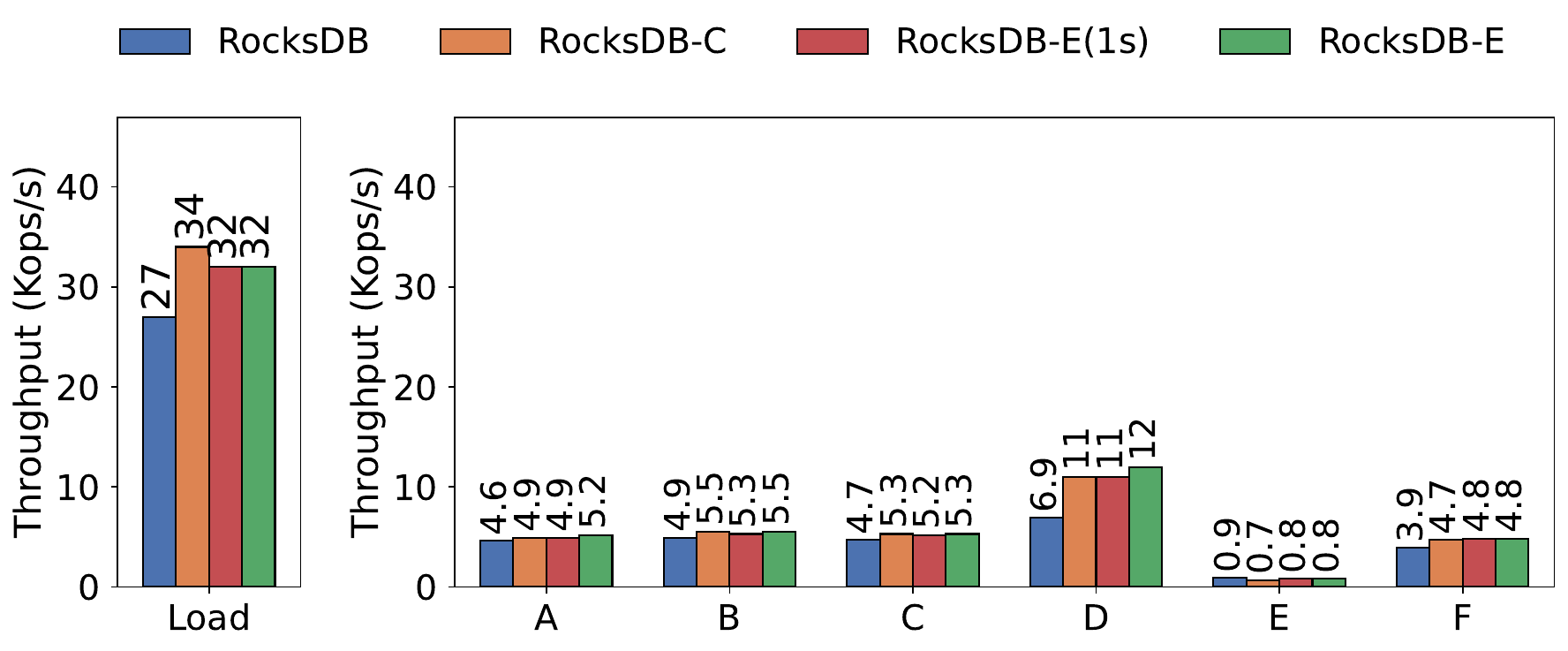}
    \caption{
    \textbf{Throughput of RocksDB across different workloads (i.e., Load and A--F) under different rate-limiting configurations.}
    These include the vanilla RocksDB without rate-limiting (i.e., \texttt{RocksDB}), RocksDB with 1-second-based IOPS limiting (i.e., \texttt{RocksDB-C}, proposed by Calcspar~\cite{zhou2023calcspar}), our proposed RocksDB with 20ms-based bandwidth-IOPS dual limiting (i.e., \texttt{RocksDB-E}), and our proposed RocksDB-E but with 1-second token refilled intervals (i.e., \texttt{RocksDB-E(1s)}) for breakdown analysis.
    }
    \label{fig:rdb_throughput}
\end{figure}

\noindent\textbf{Throughput.}
Figure~\ref{fig:rdb_throughput} presents the throughput results across different YCSB workloads (i.e., Load and A--F).
We can first see that in YCSB Load (insert-only), all three rate-limited schemes outperform the vanilla \texttt{RocksDB}. Specifically, both \texttt{RocksDB-E} variants reach 32Kops/s, which is 18.5\% higher compared to \texttt{RocksDB} without rate-limiting.
Similar trends are shown in other macro-benchmarks (i.e., A--F), where \texttt{RocksDB-E} achieves the best-level throughput, on average 20.7\% higher compared to \texttt{RocksDB}.
Although \texttt{RocksDB-C} can further improve load throughput from 32Kops/s to 34Kops/s, as discussed later, pushing ESSD beyond its bandwidth limit would cause severe latency penalties.

\begin{figure}[t]
    \centering
    \includegraphics[width=0.48\textwidth]{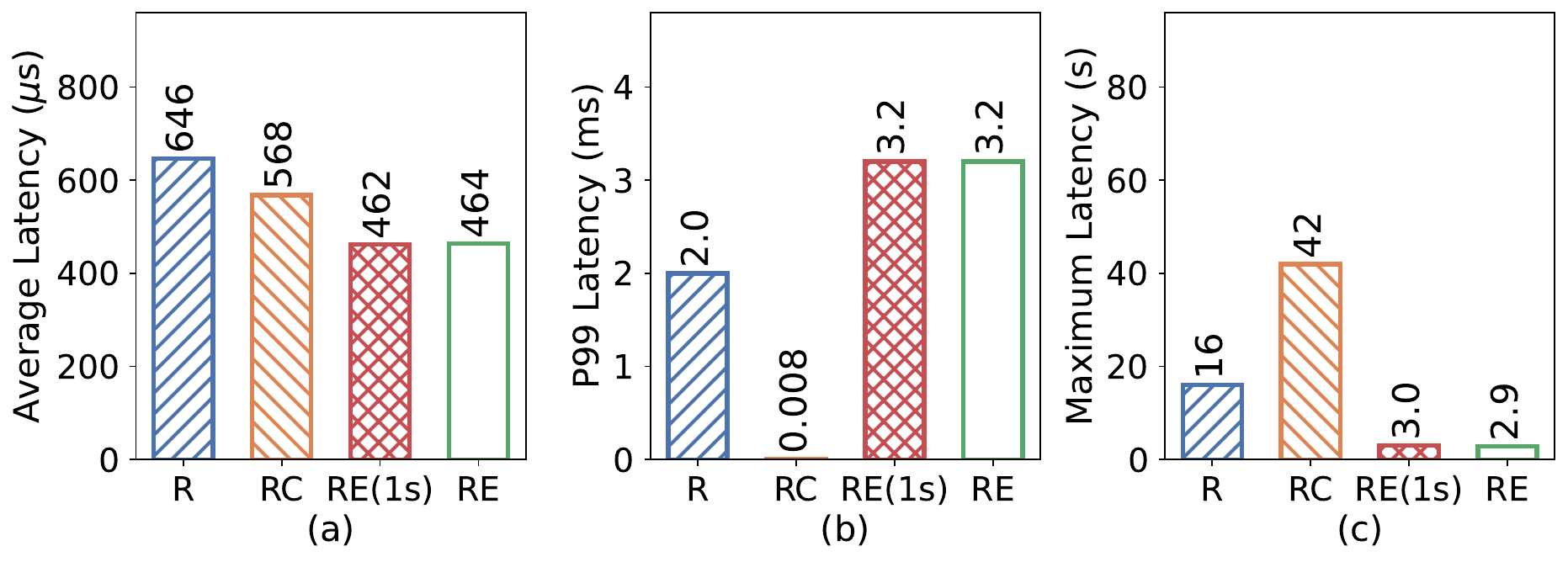}
    \caption{
    \textbf{Average, P99, and maximum latency of RocksDB in YCSB Load (insert-only) under different rate-limiting configurations.}
    \textbf{R:} \texttt{RocksDB}. 
    \textbf{RC:} \texttt{RocksDB-C}. 
    \textbf{RE(1s):} \texttt{RocksDB-E(1s)}.
    \textbf{RE:} \texttt{RocksDB-E}.
    The detailed configurations of these schemes are described in the caption of Figure~\ref{fig:rdb_throughput}.
    Note that the three subfigures use different y-axis units (i.e., $\mu$s, ms, and s).
    }
    \label{fig:rdb_load_latency}
\end{figure}

\noindent\textbf{Load Latency.}
Figure~\ref{fig:rdb_load_latency} reports the average, P99, and maximum latency results under YCSB Load.
We can first see from (a) that \texttt{RocksDB-E} reduces the average latency by 28.2\% and 18.3\% compared to \texttt{RocksDB} and \texttt{RocksDB-C}, respectively.
The results regarding maximum latency in (c) are even more striking: \texttt{RocksDB-E} reduces the maximum latency to only 2.9s, while \texttt{RocksDB} and \texttt{RocksDB-C} have maximum latencies of as high as 16s and 42s, respectively, which are 5.5$\times$ and 14.5$\times$ higher than that of \texttt{RocksDB-E}.
Given that YCSB Load is bandwidth-intensive, the disastrous performance of \texttt{RocksDB-C} reveals that bandwidth limiting (not just IOPS) is crucial.


Although the P99 latency of \texttt{RocksDB-E} in (b) is higher than both \texttt{RocksDB} and \texttt{RocksDB-C}, this reflects a deliberate architectural choice to smooth out tail latency. 
By precisely distributing I/O permits smoothly in time, \texttt{RocksDB-E} trades a modest P99 increase for the elimination of catastrophic tail events.
For SLA-governed cloud database services, effectively bounding the worst scenario to small-digit seconds is far more valuable than minimizing the P99 at the cost of encountering unpredictable outliers lasting up to 42s.

\begin{figure}[t]
    \centering
    \includegraphics[width=0.46\textwidth]{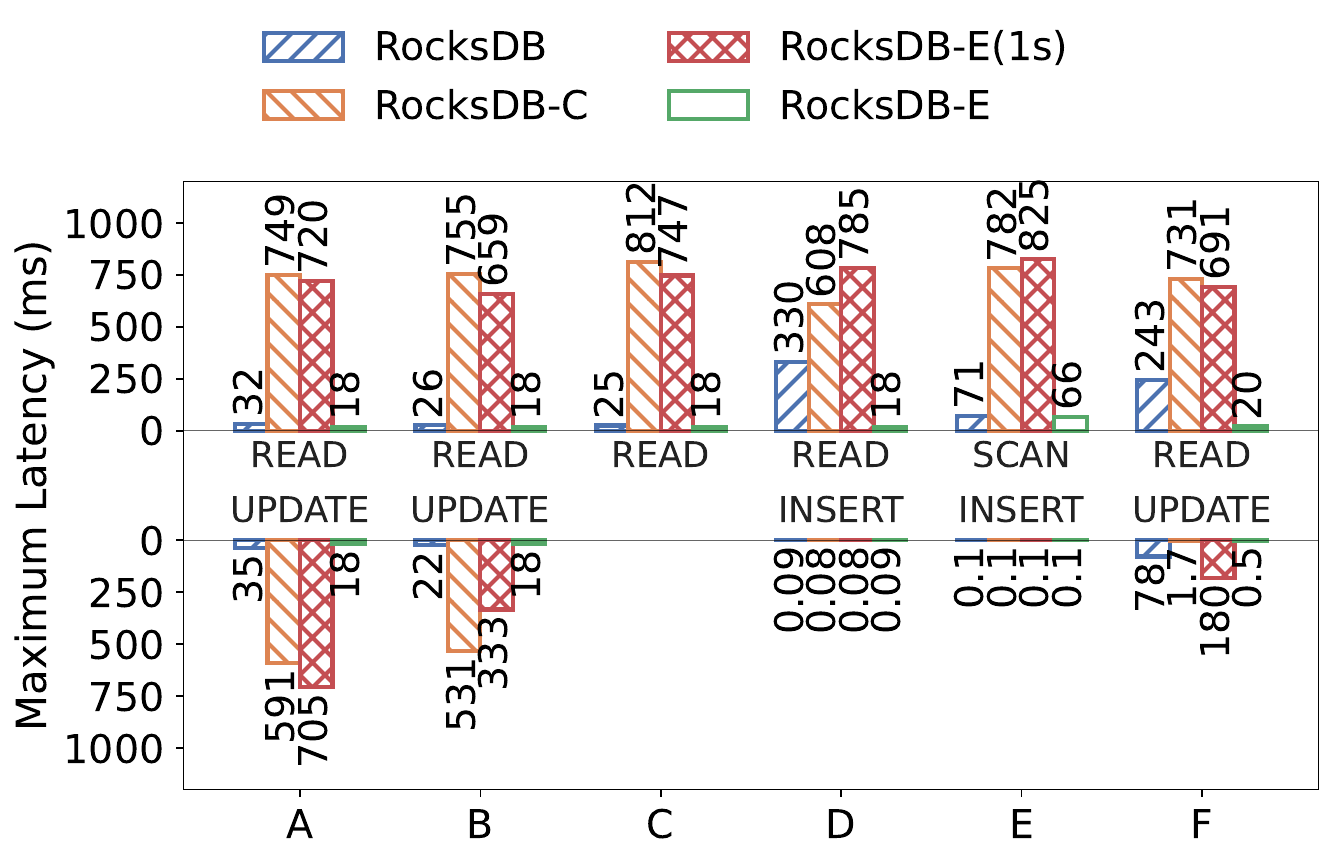}
    \caption{
    \textbf{Maximum latency of RocksDB across different macro-benchmarks (i.e., A-F) under different rate-limiting configurations.}
    The detailed configurations of these schemes are described in the caption of Figure~\ref{fig:rdb_throughput}.
    For each workload, latency is reported separately for each operation type.
    }
    \label{fig:rdb_run_lat_max}
\end{figure}

\begin{figure}[t]
    \centering
    \includegraphics[width=0.46\textwidth]{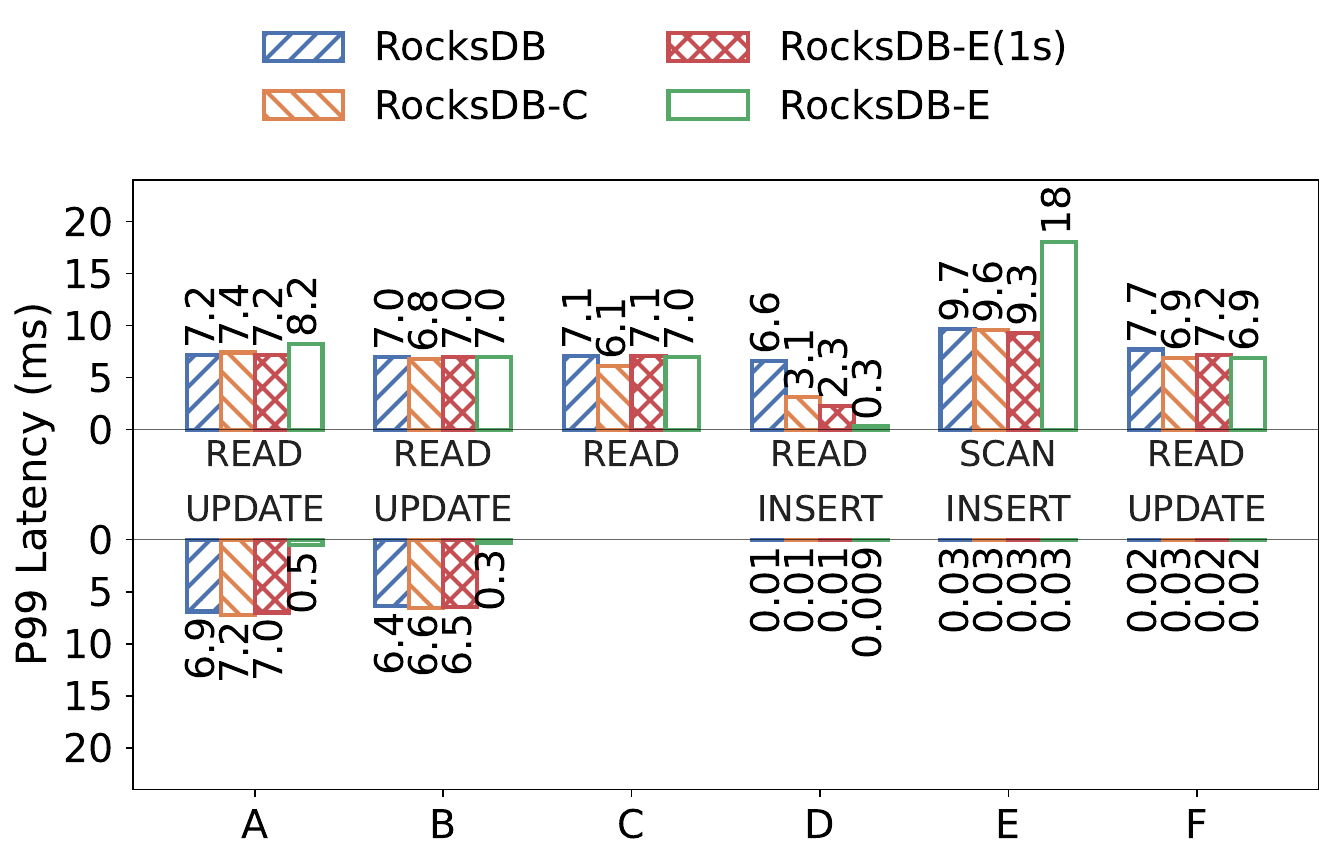}
    \caption{
    \textbf{P99 latency of RocksDB across different macro-benchmarks (i.e., A-F) under different rate-limiting configurations.}
    The detailed configurations of these schemes are described in the caption of Figure~\ref{fig:rdb_throughput}.
    For each workload, latency is reported separately for each operation type.
    }
    \label{fig:rdb_run_lat_p99}
\end{figure}

\begin{figure}[t]
    \centering
    \includegraphics[width=0.46\textwidth]{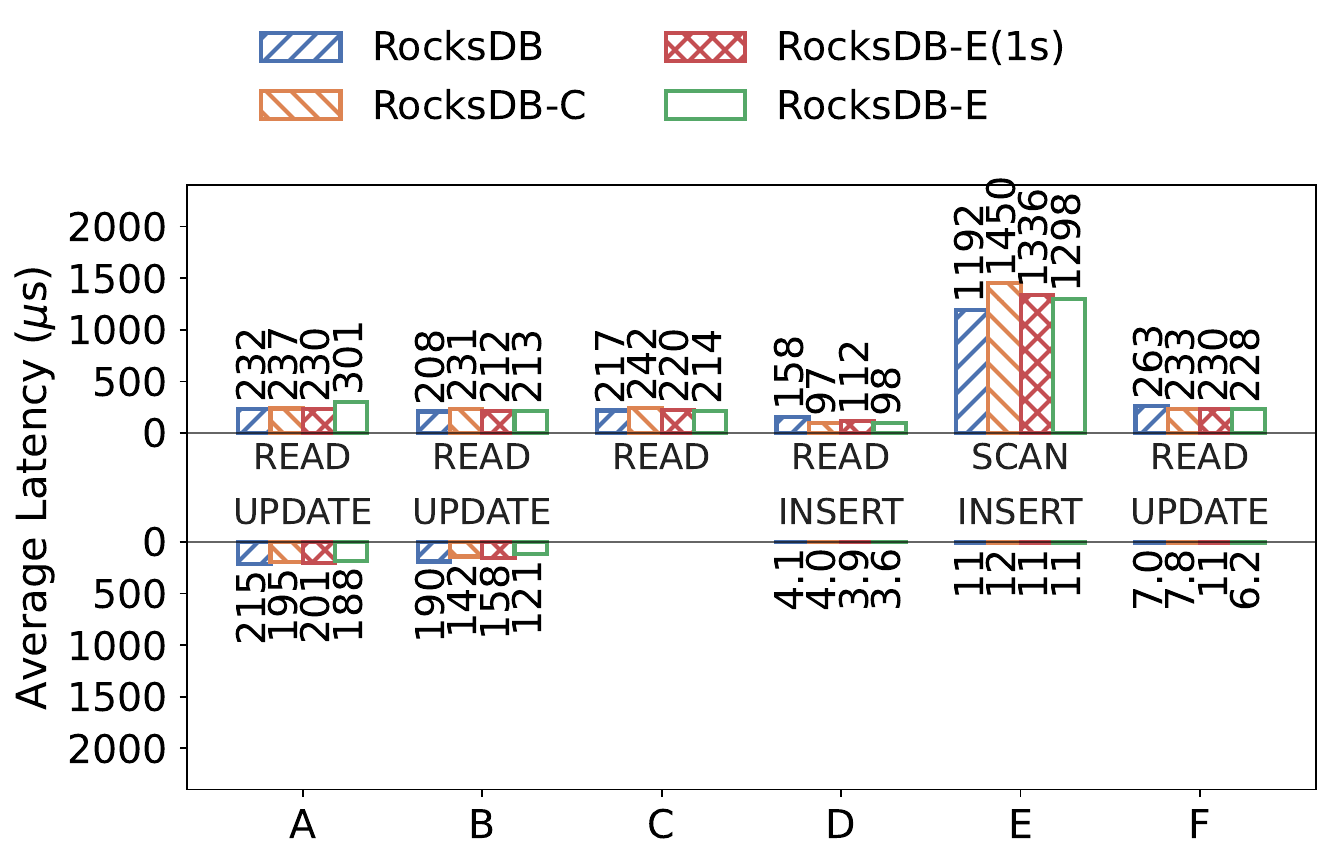}
    \caption{
    \textbf{Average latency of RocksDB across different macro-benchmarks (i.e., A-F) under different rate-limiting configurations.}
    The detailed configurations of these schemes are described in the caption of Figure~\ref{fig:rdb_throughput}.
    For each workload, latency is reported separately for each operation type.
    }
    \label{fig:rdb_run_lat_avg}
\end{figure}

\noindent\textbf{Latency in macro-benchmarks.}
Figures~\ref{fig:rdb_run_lat_max},~\ref{fig:rdb_run_lat_p99}, and~\ref{fig:rdb_run_lat_avg} report the maximum, P99, and average latency results in YCSB macro-benchmarks (i.e., A--F).
We can first see from Figure~\ref{fig:rdb_run_lat_max} that \texttt{RocksDB-E} also effectively suppresses the maximum latency in macro-benchmarks.
For example, the point read latency of \texttt{RocksDB-E} across different macro-benchmarks consistently remains between 18--20ms, and the scan latency (i.e., in YCSB E) is only 66ms.
In contrast, the maximum read latency of \texttt{RocksDB-C} ranges from 608--812ms, while that of \texttt{RocksDB-E(1s)} ranges from 659--825ms, both over an order of magnitude higher than \texttt{RocksDB-E}.
The fact that \texttt{RocksDB-E(1s)} also exhibits high maximum latency, comparable to \texttt{RocksDB-C}, demonstrates that fine-grained token refilling is crucial for securing consistently low worst-case latency in macro-benchmarks.

Regarding P99 and average latencies shown in Figures~\ref{fig:rdb_run_lat_p99} and~\ref{fig:rdb_run_lat_avg}, these two metrics follow a consistent pattern but reveal discrepancies across different operations.
Specifically, for insert and update operations across macro-benchmarks, \texttt{RocksDB-E} achieves significant improvements in both P99 latency and average latency.
However, in terms of most read and scan operations (with the exception of reading the latest data in YCSB D), the latencies of \texttt{RocksDB-E} are on par with or slightly higher than the best among all evaluated schemes.

\subsection{Revisiting Capacity and Performance Headroom (Implication \#2)}\label{sec:cs_ratelimit}

On local SSDs, garbage collection (GC) at high storage utilization causes severe write amplification and latency spikes.
High performance utilization can further exacerbate this issue, since foreground I/Os compete for the available bandwidth and IOPS, leaving less headroom for background GCs.
To this end, cloud database providers often conservatively provision storage capacity and performance budgets to secure consistent user experiences.
This approach, however, directly inflates deployment costs, as a substantial portion of paid resources remains deliberately unused.

Implication \#2 suggests an opportunity for cloud databases to reassess, rather than automatically inherit, the conservative headroom used in local-SSD deployments.
Compared to the conventional conservative approach, this directly yields more usable resources from the same hardware configurations, reducing infrastructure costs when serving the same set of tenants.

\subsection{Write Pattern: An Architectural Mismatch (Implication \#3)} \label{sec:cs_pattern}

Implication~\#3 suggests that sequential-write-based software could benefit from proactively issuing random writes, given the random write throughput advantage observed at small I/O sizes and high queue depths on high-end ESSDs.
However, we found that RocksDB is architecturally incapable of exploiting this opportunity.
The mismatch arises at two levels.

At the storage engine (e.g., LSM tree) level, all major write paths in RocksDB produce large and sequential I/Os that fall well outside the regime where random writes outperform.
Flushes and compactions generate SSTables in 1MB write batches by default.
Even for the write-ahead log (WAL) enabled with the synchronization option, its group commit technique~\cite{website:groupcommit} merges data from multiple threads into large writes.
In these cases, the I/O size far exceeds the threshold below which random writes exhibit a throughput advantage on ESSDs.

While at the filesystem level, modern general-purpose filesystems also tend to avoid random I/Os.
In ext4~\cite{mathur2007new}, extent-based allocation together with the multi-block allocator promotes contiguous block placement.
F2FS~\cite{lee2015f2fs}, by its log-structured design, converts updates into out-of-place sequential appends within active segments under its default policy.
Under both filesystems, the I/O pattern arriving at the ESSD is predominantly sequential in common deployments, leaving little room to exploit the random-write advantage.

Together, the two-level mismatch makes it difficult to apply Implication \#3 to RocksDB; otherwise, it would require invasive modifications to both mature software systems, which would be virtually impractical.
Nonetheless, purpose-built architectures that can fully control I/O patterns towards ESSDs can exploit this opportunity.
For example, SPDK-based storage engines that employ the Forced Unit Access (FUA) mechanism for strong write durability can deliberately issue small writes to non-contiguous device locations, so as to leverage the performance advantage demonstrated in Observation \#3.


\subsection{Aggressive Compression (Implication \#5)} \label{sec:cs_compress}

\begin{figure}[t]
    \centering
    \subfigure[YCSB Load (insert-only)]{
    \includegraphics[width=0.48\textwidth]{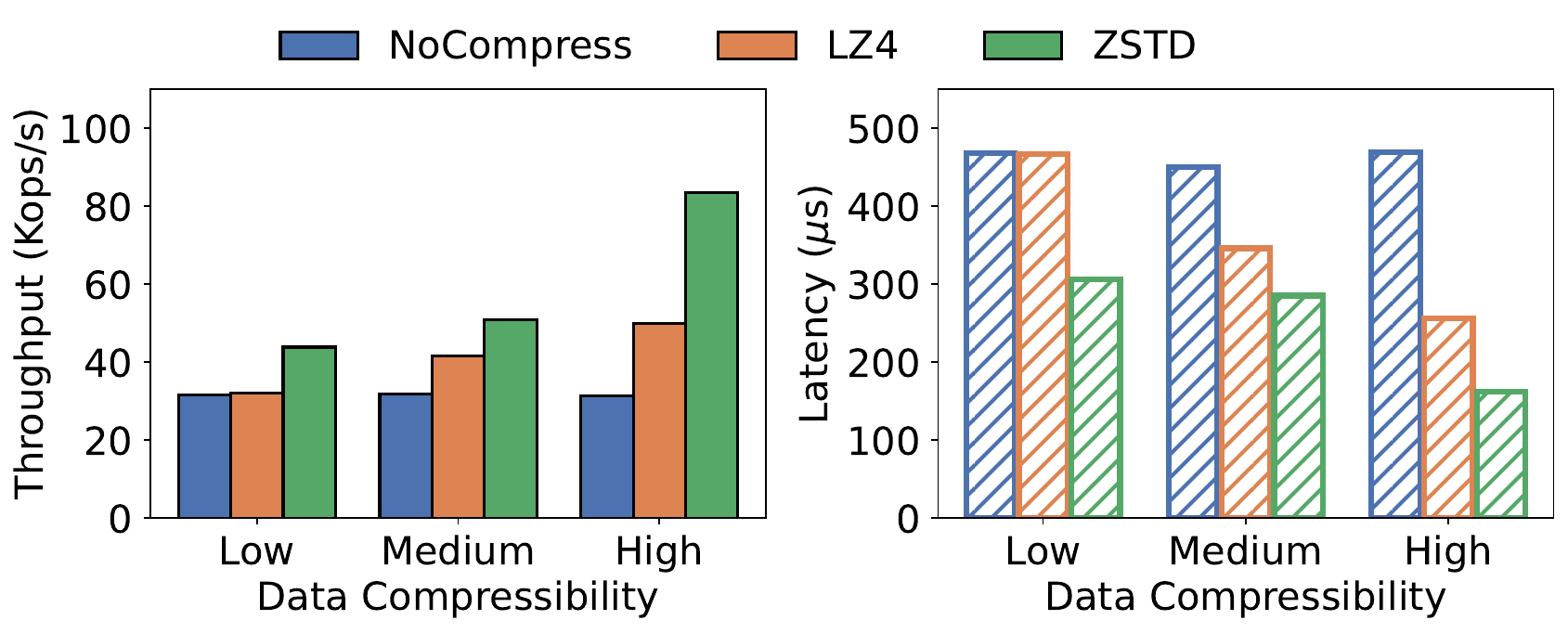}
    \label{fig:rdb_compress_insert}
    }
    \subfigure[YCSB C (read-only)]{
    \includegraphics[width=0.48\textwidth]{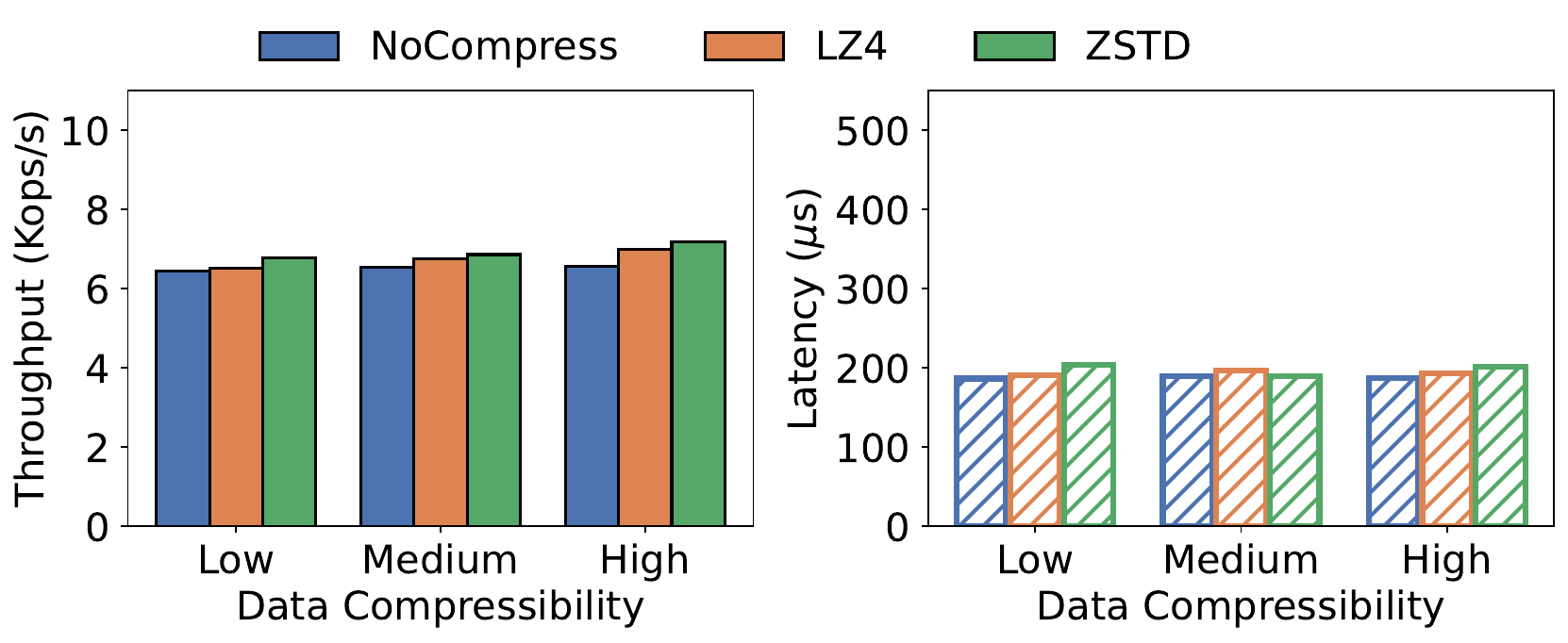}
    \label{fig:rdb_compress_read}
    }
    \subfigure[YCSB A (50\% reads and 50\% updates)]{
    \includegraphics[width=0.48\textwidth]{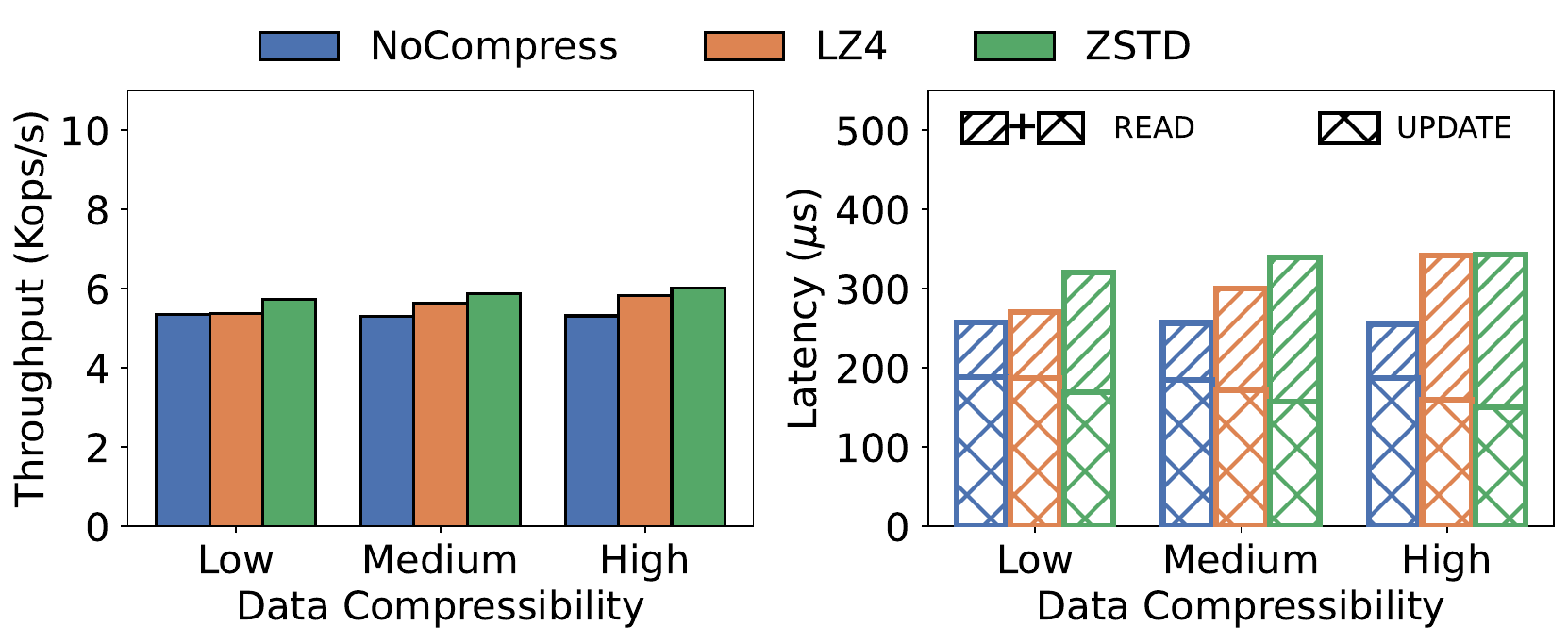}
    \label{fig:rdb_compress_a}
    }
    \caption{
    \textbf{Throughput and average latency of RocksDB across different YCSB workloads (i.e., Load, A, and C) under different compression configurations (i.e., no compression, LZ4, and ZSTD).}
    In workloads A and C, read operations follow the Zipfian distribution.
    }
    \label{fig:rdb_compress}
\end{figure}

According to Implication~\#5, we explore whether more aggressive compression using standard algorithms can benefit RocksDB on ESSDs.
We compare three compression configurations: no compression, LZ4~\cite{website:lz4}, and ZSTD~\cite{website:zstd}, across three datasets with different levels of compressibility (i.e., low, medium, and high).
On these three datasets, LZ4 achieves compression ratios of 1.01, 1.36, and 1.89, while ZSTD achieves those of 1.54, 2.11, and 3.74.
Here, compression ratio is defined as the ratio of the original size to the compressed size.
Figure~\ref{fig:rdb_compress} presents the results in terms of throughput and latency.

\noindent\textbf{Load performance.}
As shown in Figure~\ref{fig:rdb_compress_insert}, \texttt{ZSTD} dramatically improves the throughput in YCSB Load (insert-only) by 38.9\%--167\% compared to \texttt{NoCompress} and 36.8\%--67.2\% compared to \texttt{LZ4}.
Correspondingly, \texttt{ZSTD} reduces the load latency by 34.6\%--65.5\% compared to \texttt{NoCompress} and 17.6\%--36.7\% compared to \texttt{LZ4}.
The high per-I/O cost of ESSDs underscores the benefits of reducing I/O volume; consequently, aggressive compression delivers significantly better write throughput and latency, and the improvements are more pronounced with higher data compressibility.

\noindent\textbf{Read-only workload performance.}
Figure~\ref{fig:rdb_compress_read} shows the results in YCSB C (read-only).
As we can see, \texttt{ZSTD} increases the throughput by up to 9.4\% compared to \texttt{NoCompress} and up to 4.1\% compared to \texttt{LZ4}, while slightly worsening the latency by up to 8.6\% compared to \texttt{NoCompress} and up to 6.3\% compared to \texttt{LZ4}.
Here, the small performance differences are due to the high block cache hit rate, as the Zipfian access pattern concentrates reads on a hot subset that is largely served from block cache (without triggering decompression).

\noindent\textbf{Mixed read-write workload performance.}
Figure~\ref{fig:rdb_compress_a} shows the results in YCSB A (50\% reads and 50\% updates).
\texttt{ZSTD} improves the overall throughput by 6.9\%--13.3\% compared to \texttt{NoCompress} and 3.4\%--6.7\% compared to \texttt{LZ4}.
The update latency of \texttt{ZSTD} is also reduced by 10.1\%--19.8\% compared to \texttt{NoCompress} and 6.3\%--10.1\% compared to \texttt{LZ4}.
However, the read latency rises, as update-triggered compactions invalidate block cache entries, causing subsequent reads to incur decompression overhead.
Nevertheless, the read-latency gap between \texttt{ZSTD} and \texttt{LZ4} is only 0.3\%--18.5\%, narrowing to just 0.3\% for highly compressible data.
That is, even with higher decompression overhead, \texttt{ZSTD} still delivers better overall throughput and update latency; read latency is comparable on highly compressible datasets, and moderately increases on less compressible datasets.

\noindent\textbf{Storage footprint.}
Storage footprint reduction is one of the most direct cost benefits of standard compression algorithms on capacity-billed cloud storage.
In the three datasets with different compressibility (i.e., low, medium, and high), \texttt{ZSTD} significantly reduces the storage footprint by 35.2\%, 52.7\%, and 73.2\%, respectively.
In contrast, the savings from \texttt{LZ4} are only 0.7\%, 26.7\%, and 47.2\%.

\subsection{Summary} \label{sec:cs_summary}

In this section, we use RocksDB as a case study to examine how the ESSD contract and the refined I/O rate-limiting model can inform software design.
Based on the investigation, we summarize four ESSD-aware guidelines for RocksDB:
\begin{enumerate}[leftmargin=*]
    \item minimize cache misses to protect the critical read path from costly I/Os;
    \item holistically regulate I/Os with the refined rate-limiting model to improve latency, especially the maximum latency;
    \item revisit the conservative capacity and performance headroom inherited from local-SSD deployments, as GC-related degradation is not exposed at the user-visible interface;
    \item adopt aggressive compression using standard algorithms (e.g., ZSTD) when data is sufficiently compressible and CPU resources permit.
\end{enumerate}

Collectively, these results demonstrate the potential of ESSD-aware optimization in the evaluated RocksDB deployment without hardware changes.

\section{Related Work} \label{sec:related}

Characterizing the performance of emerging storage technologies remains a topic of enduring relevance.
Previous studies have explored the performance attributes of HDDs~\cite{schlosser2004mems}, flash-based SSDs~\cite{he2017unwritten}, Optane-based SSDs~\cite{wu2019towards}, as well as flash SSDs with novel interfaces such as Key-Value (KV)~\cite{saha2021kv} and Zoned Namespace (ZNS)~\cite{doekemeijer2023performance}.
Building on this research tradition, we present a performance analysis of the latest cloud-based ESSDs.

Several studies have documented advancements in EBS architectures across storage, networking, hardware acceleration, and security~\cite{miao2022luna,zhang2024s,wang2024ransom,shu2024burstable}.
For instance, Zhang et al.~\cite{zhang2024s,xu2025evolving} detail the design evolution of Alibaba Cloud EBS over ten years, tracing its progression from simplicity in EBS1 to high performance and space efficiency in EBS2, and more recently to minimizing network traffic amplification in EBS3.
Miao et al.~\cite{miao2022luna} introduce two generations of storage network stacks: Luna (a user-space TCP stack) and Solar (a storage-optimized UDP stack) developed for Alibaba Cloud EBS, which collectively reduce average I/O latency by 72\% over five years.
Shu et al.~\cite{shu2024burstable} identify performance variability in Alibaba Cloud EBS resulting from DPU-based storage agents and propose BurstCBS, a hardware-software co-designed I/O scheduler that mitigates load imbalance and tenant interference.
To address ransomware threats in EBS, Wang et al.~\cite{wang2024ransom} develop DeftPunk, a block-level detection and recovery system that attains near-perfect recall across 13 ransomware variants with minimal runtime overhead.
Additionally, Li et al.~\cite{li2023depth} profile EBS workloads from Alibaba Cloud and Tencent Cloud, offering insights into load intensity, spatial locality, and temporal patterns.
Wu et al.~\cite{wu2025hey} analyze I/O traces in Alibaba Cloud EBS, exposing traffic skew across virtualization frameworks, throttling mechanisms, cluster management, and caching layers, alongside discussions of potential mitigations.
Tan et al.~\cite{tan2025tela} propose Tela, a temporal load-aware placement strategy for ESSDs that forecasts workload patterns to enhance cloud resource allocation.

From the cloud software perspective, Cosine~\cite{chatterjee2021cosine} employs cost models to select optimal key-value storage engines based on workload requirements, budget constraints, performance targets, and cloud SLAs.
A number of studies~\cite{yoon2018mutant,xu2022building,wang2023mirrorkv} examine tiered data management using LSM trees across EBS and other storage tiers, such as faster local SSDs or slower object stores.
Mutant~\cite{yoon2018mutant} dynamically reorganizes SSTables across tiers according to access frequency, enabling adaptive cost-performance trade-offs in RocksDB without costly data migration.
ROCKSMASH~\cite{xu2022building} enhances metadata efficiency and read performance through a new cache design and extends the write-ahead log for rapid parallel recovery.
MirrorKV~\cite{wang2023mirrorkv} targets hybrid storage architectures combining EBS and object storage, introducing optimizations for compaction and query efficiency in such environments.
Finally, Zhou et al.~\cite{zhou2023calcspar} highlight elevated latency variations in ESSDs when exceeding provisioned IOPS within one-second windows and propose I/O regulation techniques in LSM-tree-based stores to improve tail latency.

\section{Conclusion} \label{sec:conclusion}

In this paper, we present a user-centric performance characterization of ESSDs from Amazon AWS and Alibaba Cloud.
Based on our in-depth investigation, we make three main contributions: (1) an ESSD contract that summarizes how ESSDs differ from local SSDs, (2) a refined I/O rate-limiting model that effectively suppresses latency spikes on ESSDs, and (3) a case study on RocksDB that explores our findings in the context of cloud databases.
Together, they serve as a practical reference for EBS users in understanding and exploiting ESSDs, and we hope that our findings can stimulate more research on investigating ESSD performance features as well as ESSD-aware cloud software designs.



\ifCLASSOPTIONcaptionsoff
  \newpage
\fi



%



\bibliographystyle{IEEEtran}
\bibliography{uc}

@inproceedings{schlosser2004mems,
  title={MEMS-based storage devices and standard disk interfaces: A square peg in a round hole?},
  author={Schlosser, Steven W and Ganger, Gregory R},
  booktitle={FAST},
  pages={87--100},
  year={2004}
}

@inproceedings{he2017unwritten,
  title={The unwritten contract of solid state drives},
  author={He, Jun and Kannan, Sudarsun and Arpaci-Dusseau, Andrea C and Arpaci-Dusseau, Remzi H},
  booktitle={Proceedings of the twelfth European conference on computer systems},
  pages={127--144},
  year={2017}
}

@inproceedings{wu2025hey,
  title={Hey Hey, My My, Skewness Is Here to Stay: Challenges and Opportunities in Cloud Block Store Traffic},
  author={Wu, Haonan and Xu, Erci and Wang, Ligang and Hong, Yuandong and Niu, Changsheng and Shi, Bo and Zhu, Lingjun and He, Jinnian and Wu, Dong and Zhang, Weidong and others},
  booktitle={Proceedings of the Twentieth European Conference on Computer Systems},
  pages={736--752},
  year={2025}
}

@article{xu2025evolving,
  title={Evolving the Cloud Block Store with Performance, Elasticity, Availability, and Hardware Offloading},
  author={Xu, Erci and Zhang, Weidong and Wang, Qiuping and Zhang, Xiaolu and Gu, Yuesheng and Lu, Zhenwei and Ouyang, Tao and Dong, Guanqun and Peng, Wenwen and Xu, Zhe and others},
  journal={ACM Transactions on Storage},
  volume={21},
  number={2},
  pages={1--29},
  year={2025},
  publisher={ACM New York, NY}
}

@inproceedings{cooper2010benchmarking,
  title={Benchmarking cloud serving systems with YCSB},
  author={Cooper, Brian F and Silberstein, Adam and Tam, Erwin and Ramakrishnan, Raghu and Sears, Russell},
  booktitle={Proceedings of the 1st ACM symposium on Cloud computing},
  pages={143--154},
  year={2010}
}

@inproceedings{tan2025tela,
  title={Tela: A Temporal Load-Aware Cloud Virtual Disk Placement Scheme},
  author={Tan, Difan and Li, Jiawei and Wang, Hua and Li, Xiaoxiao and Liu, Wenbo and Qin, Zijin and Zhou, Ke and Xie, Ming and Tao, Mengling},
  booktitle={Proceedings of the 30th ACM International Conference on Architectural Support for Programming Languages and Operating Systems, Volume 1},
  pages={1084--1100},
  year={2025}
}

@inproceedings{wu2019towards,
  title={Towards an unwritten contract of intel optane $\{$SSD$\}$},
  author={Wu, Kan and Arpaci-Dusseau, Andrea and Arpaci-Dusseau, Remzi},
  booktitle={11th USENIX Workshop on Hot Topics in Storage and File Systems (HotStorage 19)},
  year={2019}
}

@inproceedings{yoon2018mutant,
  title={Mutant: Balancing storage cost and latency in lsm-tree data stores},
  author={Yoon, Hobin and Yang, Juncheng and Kristjansson, Sveinn Fannar and Sigurdarson, Steinn E and Vigfusson, Ymir and Gavrilovska, Ada},
  booktitle={Proceedings of the ACM Symposium on Cloud Computing},
  pages={162--173},
  year={2018}
}

@article{xu2022building,
  title={Building a fast and efficient LSM-tree store by integrating local storage with cloud storage},
  author={Xu, Peng and Zhao, Nannan and Wan, Jiguang and Liu, Wei and Chen, Shuning and Zhou, Yuanhui and Albahar, Hadeel and Liu, Hanyang and Tang, Liu and Tan, Zhihu},
  journal={ACM Transactions on Architecture and Code Optimization (TACO)},
  volume={19},
  number={3},
  pages={1--26},
  year={2022},
  publisher={ACM New York, NY}
}

@inproceedings{zhou2023calcspar,
  title={Calcspar: A $\{$Contract-Aware$\}$$\{$LSM$\}$ Store for Cloud Storage with Low Latency Spikes},
  author={Zhou, Yuanhui and Zhou, Jian and Chen, Shuning and Xu, Peng and Wu, Peng and Wang, Yanguang and Liu, Xian and Zhan, Ling and Wan, Jiguang},
  booktitle={2023 USENIX Annual Technical Conference (USENIX ATC 23)},
  pages={451--465},
  year={2023}
}

@inproceedings{li2022fantastic,
  title={Fantastic SSD internals and how to learn and use them},
  author={Li, Nanqinqin and Hao, Mingzhe and Li, Huaicheng and Lin, Xing and Emami, Tim and Gunawi, Haryadi S},
  booktitle={Proceedings of the 15th ACM International Conference on Systems and Storage},
  pages={72--84},
  year={2022}
}

@inproceedings{doekemeijer2023performance,
  title={Performance characterization of nvme flash devices with zoned namespaces (zns)},
  author={Doekemeijer, Krijn and Tehrany, Nick and Chandrasekaran, Balakrishnan and Bj{\o}rling, Matias and Trivedi, Animesh},
  booktitle={2023 IEEE International Conference on Cluster Computing (CLUSTER)},
  pages={118--131},
  year={2023},
  organization={IEEE}
}

@inproceedings{saha2021kv,
  title={KV-SSD: What Is It Good For?},
  author={Saha, Manoj P and Maruf, Adnan and Kim, Bryan S and Bhimani, Janki},
  booktitle={2021 58th ACM/IEEE Design Automation Conference (DAC)},
  pages={1105--1110},
  year={2021},
  organization={IEEE}
}

@inproceedings{wang2025unwritten,
  title={The Unwritten Contract of Cloud-based Elastic Solid-State Drives},
  author={Wang, Yingjia and Yang, Ming-Chang},
  booktitle={2025 62nd ACM/IEEE Design Automation Conference (DAC)},
  pages={1--7},
  year={2025},
  organization={IEEE}
}

@article{wang2023mirrorkv,
  title={MirrorKV: An Efficient Key-Value Store on Hybrid Cloud Storage with Balanced Performance of Compaction and Querying},
  author={Wang, Zhiqi and Shao, Zili},
  journal={Proceedings of the ACM on Management of Data},
  volume={1},
  number={4},
  pages={1--27},
  year={2023},
  publisher={ACM New York, NY, USA}
}

@misc{website:awsssd,
    title = "{Amazon AWS ESSD}",
    howpublished = "\url{https://docs.aws.amazon.com/ebs/latest/userguide/ebs-volumes.html}",
    year=2026,
    }

@misc{website:awsm6in,
    title = "{Amazon EC2 General Purpose Instance Specifications}",
    howpublished = "{\url{https://docs.aws.amazon.com/ec2/latest/instancetypes/gp.html}}",
    year = {2026},
    }

@misc{website:wfq,
    title = "{Wikipedia of Weighted Fair Queueing}",
    year = 2017,
    howpublished = "\url{https://en.wikipedia.org/wiki/Weighted_fair_queueing}",
    }

@article{wang2026making,
  title={Making LSM-Tree-based Key-Value Store Practical and Efficient for Multi-Tenant Serverless Cloud Databases},
  author={Wang, Yingjia and Gong, Caixin and Zhu, Guoyun and Wang, Sheng and Wang, Zhengheng and Liu, Huan and Shi, Junzhi and Qin, Wu and Zhang, Wei and Li, Feifei and Yang, Ming-Chang},
  journal={Proceedings of the ACM on Management of Data},
  volume={4},
  number={1},
  pages={1--25},
  note={Art. no. 53},
  month=feb,
  year={2026},
  doi={10.1145/3786667},
  publisher={ACM}
}

@article{yang2020leaper,
  title={Leaper: A learned prefetcher for cache invalidation in LSM-tree based storage engines},
  author={Yang, Lei and Wu, Hong and Zhang, Tieying and Cheng, Xuntao and Li, Feifei and Zou, Lei and Wang, Yujie and Chen, Rongyao and Wang, Jianying and Huang, Gui},
  journal={Proceedings of the VLDB Endowment},
  volume={13},
  number={12},
  pages={1976--1989},
  year={2020},
  publisher={VLDB Endowment}
}

@misc{website:lz4,
    title = {{LZ4: Extremely fast compression}},
    year = 2026,
    howpublished = {\url{https://lz4.org/}},
}

@misc{website:zstd,
    title = {{Zstandard: Real-time data compression algorithm}},
    year = 2026,
    howpublished = {\url{https://facebook.github.io/zstd/}},
}

@inproceedings{bjorling2021zns,
    title = {{ZNS: Avoiding the Block Interface Tax for Flash-based SSD}},
    author = {Bj{\o}rling, Matias and Aghayev, Abutalib and Holmberg, Hans and Ramesh, Aravind and Le Moal, Damien and Ganger, Gregory R and Amvrosiadis, George},
    booktitle = {Proceedings of the {USENIX} Annual Technical Conference ({ATC’21})},
    pages = {689--703},
    year = {2021}
    }

@inproceedings{mathur2007new,
  title={The new ext4 filesystem: current status and future plans},
  author={Mathur, Avantika and Cao, Mingming and Bhattacharya, Suparna and Dilger, Andreas and Tomas, Alex and Vivier, Laurent},
  booktitle={Proceedings of the Linux symposium},
  volume={2},
  pages={21--33},
  year={2007}
}

@inproceedings{sabol2023fdp,
    title={{Flexible Data Placement: State of the Union}},
    author={Sabol, Chris and Stenfort, Ross and Allison, Mike},
    booktitle={Flash Memory Summit},
    year={2023}
}

@misc{website:iouring,
    title = "{Efficient I/O with io\_uring}",
    howpublished = "\url{https://kernel.dk/io_uring.pdf}",
    year=2019,
    }

@misc{website:tokenbucket,
    title = "{Wikipedia of Token Bucket Algorithm}",
    year = 2017,
    howpublished = "\url{https://en.wikipedia.org/wiki/Token_bucket}",
    }

@article{chatterjee2021cosine,
  title={Cosine: a cloud-cost optimized self-designing key-value storage engine},
  author={Chatterjee, Subarna and Jagadeesan, Meena and Qin, Wilson and Idreos, Stratos},
  journal={Proceedings of the VLDB Endowment},
  volume={15},
  number={1},
  pages={112--126},
  year={2021},
  publisher={VLDB Endowment}
}

@inproceedings{jiang2021fusionraid,
  title={$\{$FusionRAID$\}$: Achieving Consistent Low Latency for Commodity $\{$SSD$\}$ Arrays},
  author={Jiang, Tianyang and Zhang, Guangyan and Huang, Zican and Ma, Xiaosong and Wei, Junyu and Li, Zhiyue and Zheng, Weimin},
  booktitle={19th USENIX Conference on File and Storage Technologies (FAST 21)},
  pages={355--370},
  year={2021}
}

@misc{website:ratelimiter,
    title = "{Rate Limiter in RocksDB}",
    year = 2022,
    howpublished = "\url{https://github.com/facebook/rocksdb/wiki/rate-limiter}",
    }

@misc{website:groupcommit,
    title = "{WAL Group Commit in RocksDB}",
    year = 2020,
    howpublished = "\url{https://github.com/facebook/rocksdb/wiki/WAL-Performance}",
    }

@misc{website:alissd,
    title = "{Alibaba Cloud ESSD}",
    howpublished = "\url{https://www.alibabacloud.com/help/en/ecs/user-guide/essds}",
    year=2026,
    }

@misc{website:rocksdb,
    title = "{RocksDB: A persistent key-value store for fast storage environments}",
    year = 2026,
    howpublished = "\url{http://rocksdb.org/}",
    }

@misc{website:samsung970pro,
    title = "{Samsung 970 Pro}",
    howpublished = "\url{https://semiconductor.samsung.com/consumer-storage/internal-ssd/970pro/}",
    year=2024,
    }

@inproceedings{lee2015f2fs,
  title={$\{$F2FS$\}$: A new file system for flash storage},
  author={Lee, Changman and Sim, Dongho and Hwang, Jooyoung and Cho, Sangyeun},
  booktitle={13th USENIX Conference on File and Storage Technologies (FAST 15)},
  pages={273--286},
  year={2015}
}

@article{rodeh2013btrfs,
  title={BTRFS: The Linux B-tree filesystem},
  author={Rodeh, Ohad and Bacik, Josef and Mason, Chris},
  journal={ACM Transactions on Storage (TOS)},
  volume={9},
  number={3},
  pages={1--32},
  year={2013},
  publisher={ACM New York, NY, USA}
}

@inproceedings{skourtis2014flash,
  title={Flash on rails: Consistent flash performance through redundancy},
  author={Skourtis, Dimitris and Achlioptas, Dimitris and Watkins, Noah and Maltzahn, Carlos and Brandt, Scott},
  booktitle={2014 USENIX Annual Technical Conference (USENIX ATC 14)},
  pages={463--474},
  year={2014}
}

@inproceedings{kim2019alleviating,
  title={Alleviating garbage collection interference through spatial separation in all flash arrays},
  author={Kim, Jaeho and Lim, Kwanghyun and Jung, Youngdon and Lee, Sungjin and Min, Changwoo and Noh, Sam H},
  booktitle={2019 USENIX Annual Technical Conference (USENIX ATC 19)},
  pages={799--812},
  year={2019}
}

@misc{website:fio,
    title = "{FIO benchmark}",        
    howpublished = "\url{https://fio.readthedocs.io/en/latest/}",
    year = 2026,
    }

@inproceedings{miao2022luna,
  title={From luna to solar: the evolutions of the compute-to-storage networks in alibaba cloud},
  author={Miao, Rui and Zhu, Lingjun and Ma, Shu and Qian, Kun and Zhuang, Shujun and Li, Bo and Cheng, Shuguang and Gao, Jiaqi and Zhuang, Yan and Zhang, Pengcheng and others},
  booktitle={Proceedings of the ACM SIGCOMM 2022 Conference},
  pages={753--766},
  year={2022}
}

@inproceedings{wang2024ransom,
  title={Ransom Access Memories: Achieving Practical Ransomware Protection in Cloud with $\{$DeftPunk$\}$},
  author={Wang, Zhongyu and Song, Yaheng and Xu, Erci and Wu, Haonan and Tong, Guangxun and Sun, Shizhuo and Li, Haoran and Liu, Jincheng and Ding, Lijun and Liu, Rong and others},
  booktitle={18th USENIX Symposium on Operating Systems Design and Implementation (OSDI 24)},
  pages={687--702},
  year={2024}
}

@inproceedings{zhang2024s,
  title={What's the Story in $\{$EBS$\}$ Glory: Evolutions and Lessons in Building Cloud Block Store},
  author={Zhang, Weidong and Xu, Erci and Wang, Qiuping and Zhang, Xiaolu and Gu, Yuesheng and Lu, Zhenwei and Ouyang, Tao and Dai, Guanqun and Peng, Wenwen and Xu, Zhe and others},
  booktitle={22nd USENIX Conference on File and Storage Technologies (FAST 24)},
  pages={277--291},
  year={2024}
}

@article{li2023depth,
  title={An in-depth comparative analysis of cloud block storage workloads: Findings and implications},
  author={Li, Jinhong and Wang, Qiuping and Lee, Patrick PC and Shi, Chao},
  journal={ACM Transactions on Storage},
  volume={19},
  number={2},
  pages={1--32},
  year={2023},
  publisher={ACM New York, NY}
}

@inproceedings{shu2024burstable,
  title={Burstable Cloud Block Storage with Data Processing Units},
  author={Shu, Junyi and Qian, Kun and Zhai, Ennan and Liu, Xuanzhe and Jin, Xin},
  booktitle={18th USENIX Symposium on Operating Systems Design and Implementation (OSDI 24)},
  pages={783--799},
  year={2024}
}

\end{document}